\documentclass[twocolumn,10pt,cites,graphicx]{article}
\usepackage{epsfig}
\usepackage{graphicx}
\usepackage{amsmath}
\usepackage[dvips]{color}

\title{Point defects in two-dimensional colloidal crystals: simulation vs. elasticity theory.}
\author{Wolfgang Lechner\\
University of Vienna, Faculty of Physics, Boltzmanngasse 5, Vienna, Austria. \\
E-mail: wolfgang.lechner@univie.ac.at \and
Christoph Dellago\\
University of Vienna, Faculty of Physics, Boltzmanngasse 5,
Vienna, Austria. \\
E-mail: christoph.dellago@univie.ac.at}

\date{Received XXXXth Month, 200X\\Accepted XXXXth Month,
200X\\DOI: 10.1039/}

\begin{document}

\maketitle
\renewcommand{\thefootnote}{\fnsymbol{footnote}}

\vspace{1cm}

\noindent Using numerical and analytical calculations we study the structure of vacancies and interstitials in two-dimensional colloidal crystals. In particular, we compare the displacement fields of the defect obtained numerically with the predictions  of continuum elasticity theory for a simple defect model. In such a comparison it is of crucial importance to employ corresponding boundary conditions both in the particle and in the continuum calculations. Here, we formulate the continuum problem in a way that makes it analogous to the electrostatics problem of finding the potential of a point charge in periodic boundary conditions. The continuum calculations can then be carried out using the technique of Ewald summation. For interstitials, the displacement fields predicted by elasticity theory are accurate at large distances, but large deviations occur near the defect for distances of up to 10 lattice spacings. For vacancies, the elasticity theory predictions obtained for the simple model do not reproduce the numerical results even far away from the defect. 


\section{Introduction}

Many properties of crystalline materials are strongly affected by the presence of imperfections in the crystal lattice. In particular, point defects such as vacancies and self interstitials have a profound influence on the mechanical, optical, and electrical behavior of the material. Recent advances in experimental techniques for the manipulation and observation of colloidal systems \cite{SCIENCE_GASSER,CONFOCAL} now permit to study the fundamental properties of point defects in condensed matter systems with ``atomistic'' space and time resolution. Using optical tweezers to manipulate individual colloidal particles, Pertsinidis and Ling \cite{PERTSINIDIS_NATURE,PERTSINIDIS_NJP,PERTSINIDIS_PRL} have generated point defects in two-dimensional crystals and have studied their stable structures, interactions and diffusion. In other experimental work, Maret, Gr{\"u}nberg and collaborators \cite{MELTING_2D,MARET,GRUENBERG_PAIR} have investigated the effective interactions of thermally excited topological defects in crystals of paramagnetic colloidal particles and discussed the significance of these interactions for 2d-melting, which according to the celebrated Kosterlitz-Thouless-Halperin-Nelson-Young theory \cite{KTHNY}, involves the formation and dissociation of topological defect pairs. Point defects also play an important role in the two-dimensional electron lattice, the so called Wigner crystal \cite{WIGNER}, in which they carry implication for the melting mechanism \cite{ANDERSENSWOPE,ANDERSENSWOPE2,ANDERSENSWOPEMETHOD}, and for the the conjectured supersolid phase of Helium 4 \cite{SUPERSOLID}, in which case the attractive interactions of vacancies and interstitial may lead to expulsion of defects from the crystal thus preventing formation of a supersolid \cite{TROYER}.

From experiments  \cite{PERTSINIDIS_NATURE,PERTSINIDIS_NJP,PERTSINIDIS_PRL} and computer simulations \cite{REICHHARDT,DASILVA} it is known that vacancies and interstitials in 2d colloidal crystals can occur in various stable configurations with symmetries that differ from the symmetry of the underlying lattice. In the present article we study the structure and energetics of such point defects in a 2d crystal of soft spheres using computer simulations and analytical calculations. In particular, we address the question of how accurately the disturbances created by point defects can be rationalized in terms of elastic continuum theory. Due to the long range nature of elastic displacement fields, in carrying out such a comparison it is critical to use corresponding boundary conditions in the particle and continuum calculation. Similar periodic image effects due to elastic interactions need to be taken into account also in the atomistic modeling  of dislocations \cite{YIP1,YIP2}. As we show below, the structure of point defects in a system with periodic boundary conditions can be determined within elasticity theory with the technique of Ewald summation familiar from the computer simulation of systems with electrostatic interactions \cite{FRENKEL_SMIT}. This technique has been used before to adapt the interaction of dislocations to periodic boundary conditions \cite{WIGNER,LADD,BLADON}. Here we use Ewald summation to solve the equilibrium condition of elasticity theory and calculate the displacement field of a simple point defect model under periodic boundary conditions. While elasticity theory accurately describes the lattice distortion caused by point defects in the far field, non-linearities and discrete lattice effects dominate the defect structure near the defect. 

Although all the numerical studies discussed in this paper are carried out for two-dimensional crystals of soft spheres, simulations performed for three-dimensional crystals of various structures and with different interaction potentials, including Gaussian core, Lennard-Jones, screened electrostatic, and $1/r^3$-interactions, indicate that the phenomena described here are common to many atomic and colloidal systems. 

The remainder of the paper is organized as follows. In Sec. \ref{sec:disp} we describe how we determine the displacement fields of point defects numerically, discuss whether such calculations should be done at constant pressure or at constant volume, and present the displacement fields caused by interstitials and vacancies in various configurations. The elasticity theory formalism we use to analyze the displacement patterns of point defects is developed in Sec. \ref{sec:elastic}. In this section, we also discuss the analogy between elasticity theory and electrostatics that enables us to use the method of Ewald summation to obtain displacement fields from elasticity theory. These displacement fields are compared to those obtained numerically in Sec. \ref{sec:compare}. A summary and conclusions are provided in Sec. \ref{sec:conclusion}.


\section{Displacement fields}  
\label{sec:disp}
 
Throughout this paper, we use the Gaussian core model as a generic model for a system of soft spheres  \cite{STILLINGER_GCM,GAUSSIAN_CORE_PHASE,GAUSSIAN_SOFTLY_PHASE}. 
In this purely repulsive system, pairs of particles interact via
\begin{equation}
v(r) = \varepsilon \exp(-r^2/\sigma^2)
\end{equation}
where $r$ is the inter-particle distance and $\varepsilon$ and $\sigma$ set the energy and length scales, respectively. In the following, energies are measured in units of $\varepsilon$ and distances in units of $\sigma$. The Gaussian core model, often studied in soft condensed matter science, accurately describes the short-ranged effective interactions between polymer coils in solution \cite{FLORY}. Depending on temperature and density, the three-dimensional Gaussian core model can exist as a fluid, a bcc- or an fcc-solid  \cite{GAUSSIAN_CORE_PHASE}. In two dimensions, the perfect triangular lattice is the lowest energy structure at all densities \cite{GCM_STILL_2d_1}. Since Gaussian core particles are purely repulsive, they can form stable crystals only at pressures larger than zero. The two-dimensional Gaussian core model, which approaches the hard disk system in the limit of low temperature and low density \cite{GCM_STILL_2d_1}, has been used previously to study the melting transition in two dimensions \cite{GCM_STILL_2d_1,GCM_STILL_2d_2}.

To make contact between numerical calculations in the particle system and continuum elasticity theory, we determine, at $T=0$, the displacement field \cite{LANDAU_LIFSCHITZ}
\begin{equation}
\label{equ:displacementfield}
{\bf u}({\bf r}_i) \equiv {\bf r}'_i - {\bf r}_i
\end{equation}
caused by the introduction of the defect into the system. Here, ${\bf r}'_i$ and ${\bf r}_i$ denote the position of particle $i$ with and without the defect, respectively. The displacemet field completely describes the response of the system's structure to the perturbation introduced by the defect. Numerically, we determine displacement fields by inserting a particle into or removing it from a perfect crystal on a triangular lattice. The system is then relaxed to a new minimum energy configuration by steepest descent minimization at constant volume of the simulation box. Periodic boundary conditions apply. Typically, about tens of thousands of steepest descent steps are required to determine minimum energy structures accurately. In each minimization step, each particle is moved in the direction of the force acting on the particle where the absolute value of the displacement in chosen to be small enough to ensure that the energy of the system is reduced in each step. The displacement ${\bf u}({\bf r}_i)$ of particle $i$ is then simply the vector which connects the position of particle $i$ before the minimization, ${\bf r}_i$, with its position after the minimization, ${\bf r}'_i$.

The largest system we study here consists of  $N=199,680$ Gaussian core particles (without the extra particle) at a number density of $\rho = 0.6 \sigma^{-2}$ corresponding to a lattice constant of $a = 1.3872 \sigma$. The almost square simulation box has length $L_x = 416 a$ and height $L_y = (\sqrt{3}/2) 480 a = 415.692 a$ with aspect ratio $L_y/L_x=0.99926$. 

\subsection{Constant $V$ or constant $p$?}

In calculating the displacement fields caused by point defects the question naturally arise whether one should do that at constant volume $V$ or at constant pressure $p$. Naturally, the choice should depend on the particular experimental situation one is interested in. As we will show here, however, the displacement fields caused by a point defect at constant pressure and at constant volume are simply related. To determine how they are related, consider a perfect triangular crystal at $T=0$ enclosed in a rectangular cell of volume $V_0$ with appropriate aspect ratio. For this particular volume, the crystal is under the hydrostatic pressure $p_0$. Insertion of a point defect into the crystal at a fixed total volume distorts the crystal and atom $i$ is displaced by
\begin{equation}
{\bf u}_{V_0}({\bf r}_i)={\bf r}'_i (V_0)-{\bf r}_i(V_0),
\end{equation}
where the subscript $V_0$ indicates that the displacement field ${\bf u}_{V_0}({\bf r}_i)$ is obtained at constant volume $V_0$. In the above equation, ${\bf r}'_i (V_0)$ and ${\bf r}_i(V_0)$ are the positions of atom $i$ in the system of volume $V_0$ with and without the defect, respectively. If one requires, however, that the defects is created at constant pressure $p_0$, the volume of the simulation cell changes from $V_0$ to $V_1$ (typically, it will increase for an interstitial and decrease for a vacancy) and the atoms are displaced by a different amount,
\begin{equation}
{\bf u}_{p_0}({\bf r}_i)={\bf r}'_i (V_1)-{\bf r}_i(V_0),
\end{equation}
where the subscript $p_0$ implies that the displacement field is considered at constant pressure. Note that here we assume that during the generation of the defect the simulations cell only expands or contracts, but does not change its shape. This assumption can be lifted as discussed below. We now imagine that the defect generation at constant pressure is carried out in two steps: first the system is homogeneously dilated without defect from volume $V_0$ to volume $V_1$; in the second step, the defect is inserted at constant volume $V_1$. This two step operation corresponds to adding and subtracting ${\bf r}_i(V_1)$, i.e., the position of atom $i$ at volume $V_1$ in the absence of the defect, to the right hand side of the above equation,
\begin{equation}
{\bf u}_{p_0}({\bf r}_i)={\bf r}_i (V_1)-{\bf r}_i(V_1)+{\bf r}'_i (V_1)-{\bf r}_i(V_0).
\end{equation}
What one obtains in this way is
\begin{equation}
{\bf u}_{p_0}({\bf r}_i)={\bf u}_{V_1}({\bf r}_i)+{\bf u}_h({\bf r}_i, V_0, V_1),
\end{equation}
where ${\bf u}_{V_1}({\bf r}_i)={\bf r}'_i (V_1)-{\bf r}_i(V_1)$ is the displacement field obtained by inserting the defects at volume $V_1$ for fixed simulation cell and ${\bf u}_h({\bf r}_i, V_0, V_1)={\bf r}_i (V_1)-{\bf r}_i(V_0)$ is the displacement field corresponding to a homogeneous dilatation (or contraction) of the perfect crystal without defect from volume $V_0$ to volume $V_1$. This simple deformation corresponds to a displacement  ${\bf u}_h({\bf r}_i, V_0, V_1)=(V_1/V_0)^{1/2} {\bf r}_i$ (in three dimensions the exponent is $1/3$). Hence the displacement fields for constant pressure and constant volume are related by:
\begin{equation}
{\bf u}_{p_0}({\bf r}_i)={\bf u}_{V_1}({\bf r}_i)+\sqrt{\frac{V_1}{V_0}} {\bf r}_i.
\end{equation}
Thus, one can determine the constant-pressure displacement at pressure $p_0$ by calculating the constant-volume displacement at volume $V_1$, the volume at pressure $p_0$ in the presence of the defect.   

Similar considerations can be used to relate the constant-pressure and constant-volume displacement fields if the simulation cell is permitted to change shape as well as volume during the constant-pressure defect insertion. In this case, the simulation cell is characterized by the vectors ${\bf a}$ and ${\bf b}$ along its edges \cite{RAHMAN_PARRINELLO}. It then turns out that the displacement field of a defect inserted into an initially rectangular simulation cell with edge vectors ${\bf a}$ and ${\bf b}$ at constant hydrostatic pressure is simply related to the displacement field for fixed cell vectors ${\bf a}'$ and ${\bf b}'$ which, in general, differ from ${\bf a}$ and ${\bf b}$. For sufficiently large systems, however, a fixed shape of the simulation cell is only a very weak constraint. In particular, a displacement field which tends to be isotropic at large distances may lead to a change in aspect ratio of the simulation cell at constant pressure, but not to a change in the relative orientation of the edge vectors. All calculations of this paper are carried out for fixed and nearly square simulation cells. 

\subsection{Interstitials}

\begin{figure}[htb,floatfix]
\centerline{\includegraphics[width=7.5cm]{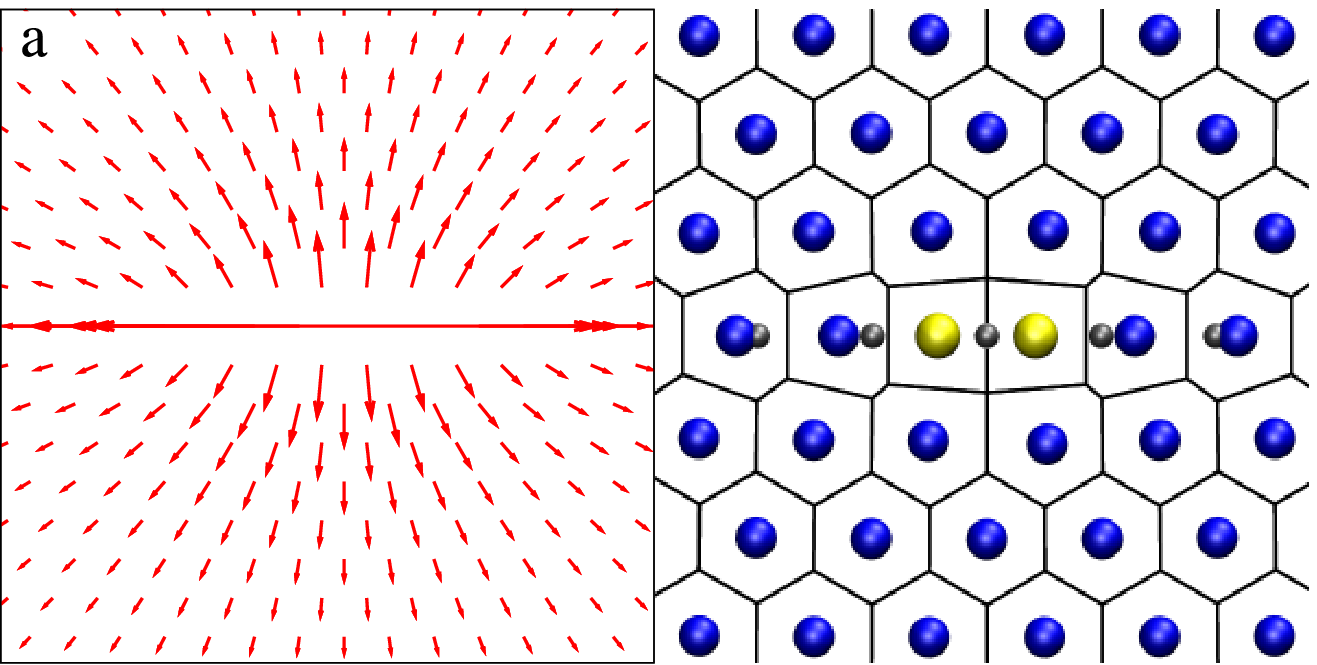}}
\vspace{0.5cm}
\centerline{\includegraphics[width=7.5cm]{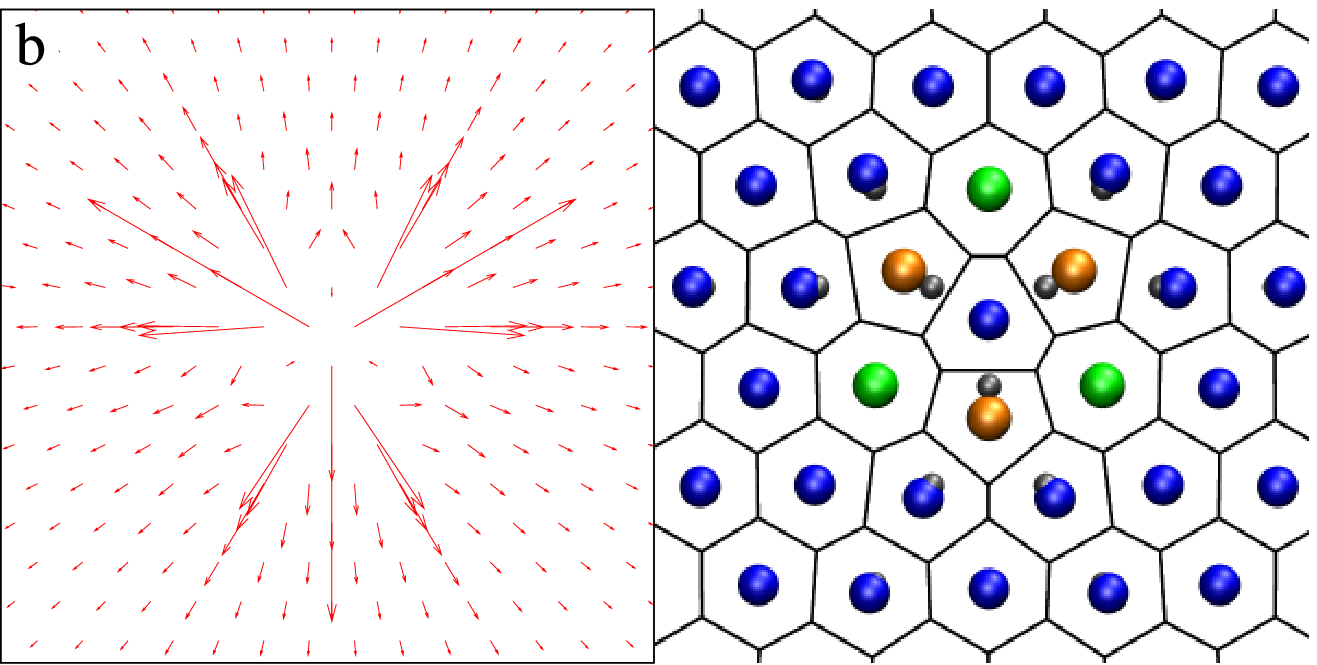}}
\vspace{0.5cm}
\centerline{\includegraphics[width=7.5cm]{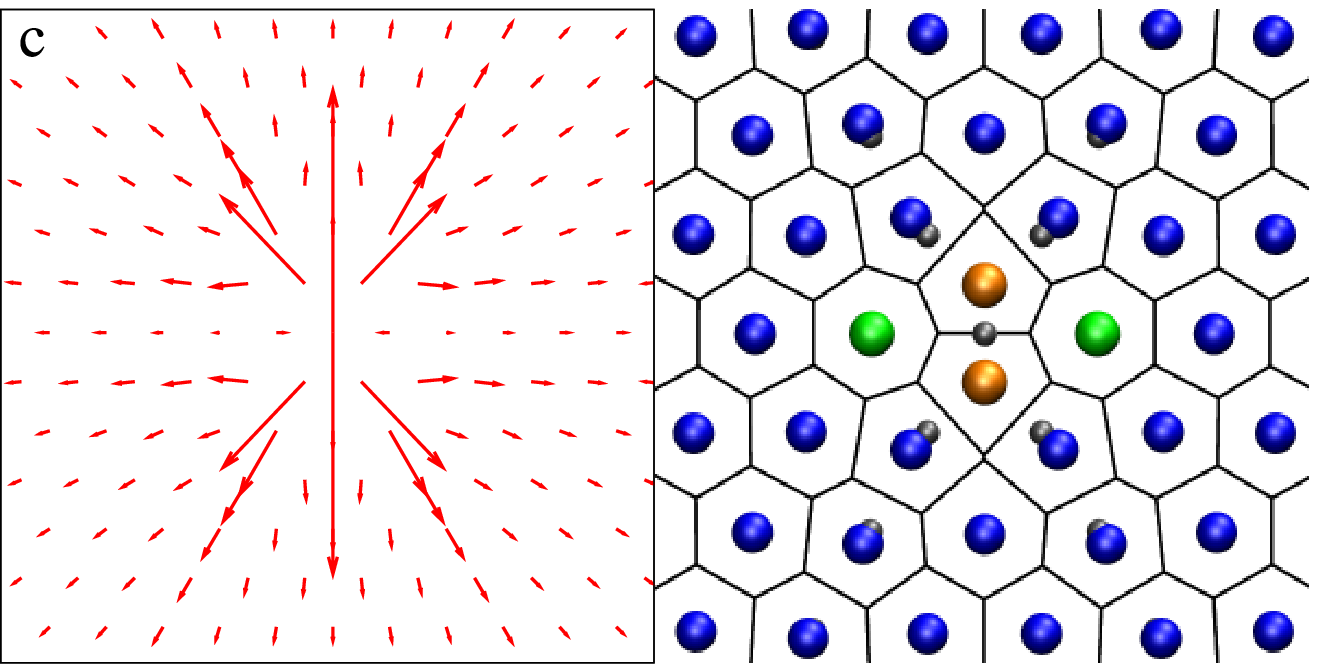}}
\caption{Displacement fields (left hand side) and particle configurations (right hand side) of the $I_2$ interstitial (a), the $I_3$ interstitial (b) and the $I_d$ interstitial (c). The arrows representing the particle displacements are exaggerated in length by a factor of 20 for better visibility. On the right hand side, the small grey spheres indicate the sites of the perfect triangular lattice. The blue spheres represent particles with $6$ neighbors according to the Voronoi construction (black lines). Yellow spheres are particles with $4$ neighbors, orange and green spheres represent particles with $5$ and $7$ neighbors, respectively. }
\label{fig:displacementinter}
\end{figure}

We first determine the displacement field of a single interstitial particle. This type of point defect can exist in different configurations \cite{PERTSINIDIS_NJP} with displacement fields of different symmetries \cite{CODEF}. The three lowest energy structures are shown in Fig. \ref{fig:displacementinter}. In one minimum-energy configuration, termed $I_2$ interstitial and shown in Fig. \ref{fig:displacementinter}a, the extra particle and one of the original particles arrange themselves at equal distance around the lattice position of the original particle leading to a two-fold symmetry. This is the two dimensional analogue of the crowdion in an fcc crystal \cite{CROWDION}. The displacements are largest on the main defect axis, which can be aligned in any of the three low-index directions of the lattice. Since the defect symmetry differs from that of the underlying triangular lattice, one may wonder whether the rectangular periodic boundary conditions used in the calculation favor the two-fold defect symmetry. Calculations carried out with hexagonal boundary conditions, however, yield identical results demonstrating that the defect symmetry is not imposed by the symmetry of the boundary conditions.

Other low-energy defect configurations include the $I_3$ interstitial with three-fold symmetry shown in Fig. \ref{fig:displacementinter}b and the $I_d$ interstitial or dumbbell interstitial shown in Fig. \ref{fig:displacementinter}c. In the $I_3$ configuration, the extra particle is located at the center of a triangle spanned by three nearest neighbor lattice points and the surrounding particles are displaced outward with respect to their original positions.  In the dumbbell configuration, the interstitial particle and one of the original particles compete for one lattice position as in the $I_2$ interstitial, but the line connecting them is orthogonal to one of the low-index lattice directions.  In contrast to the $I_2$ interstitial, the $I_d$ is not concentrated on one single axis. 

The interstitial configurations observed in the Gaussian core model have energies that differ by less than 0.1\% of the total defect energy. These energy difference correspond to roughly 20\% of the thermal energy $k_{\rm B}T$ at melting. At finite temperatures that are not too low, interconversion between the various defect configurations is facile and all three of them play an important role during defect diffusion  \cite{REICHHARDT}. 

\subsection{Vacancies}

\begin{figure}[htb,floatfix]
\centerline{\includegraphics[width=7.5cm]{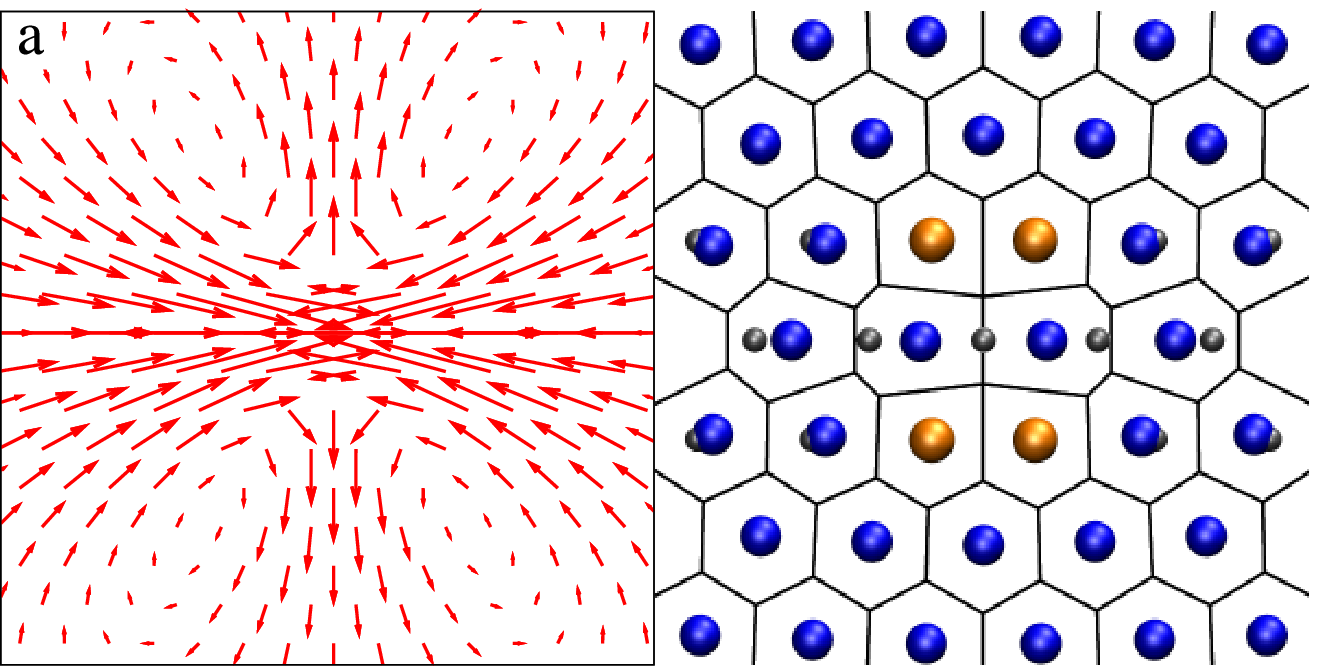}}
\vspace{0.5cm}
\centerline{\includegraphics[width=7.5cm]{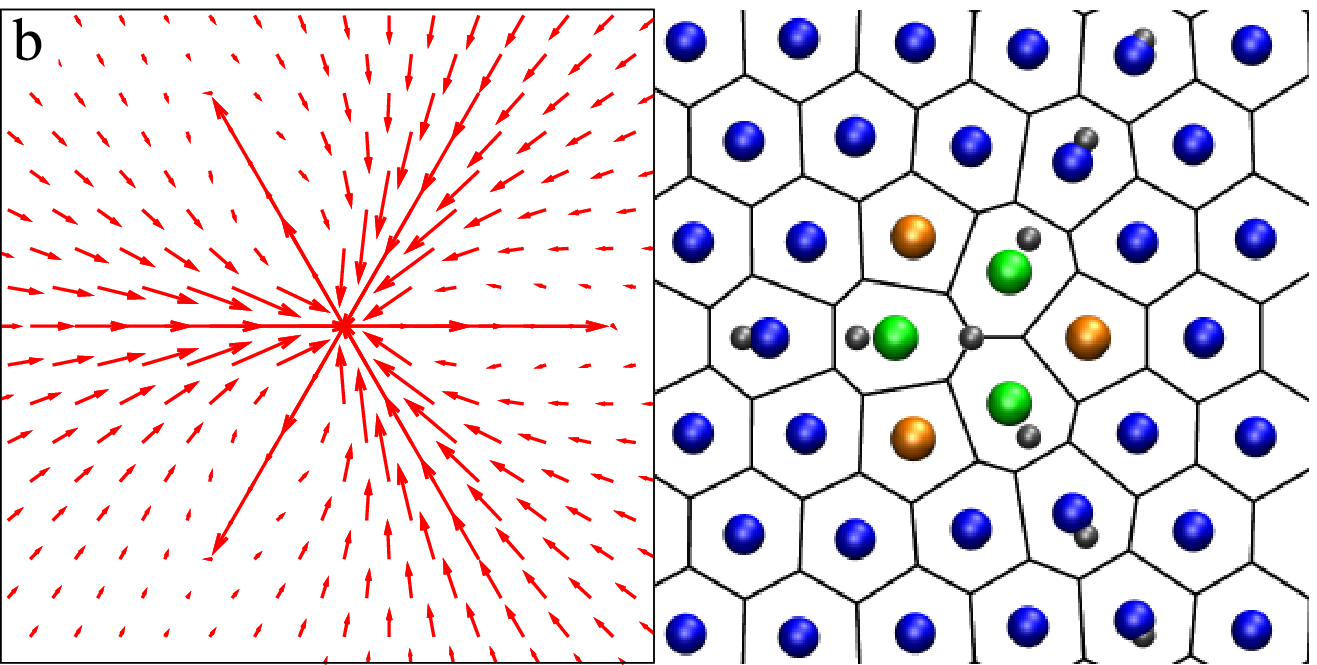}}
\vspace{0.5cm}
\centerline{\includegraphics[width=7.5cm]{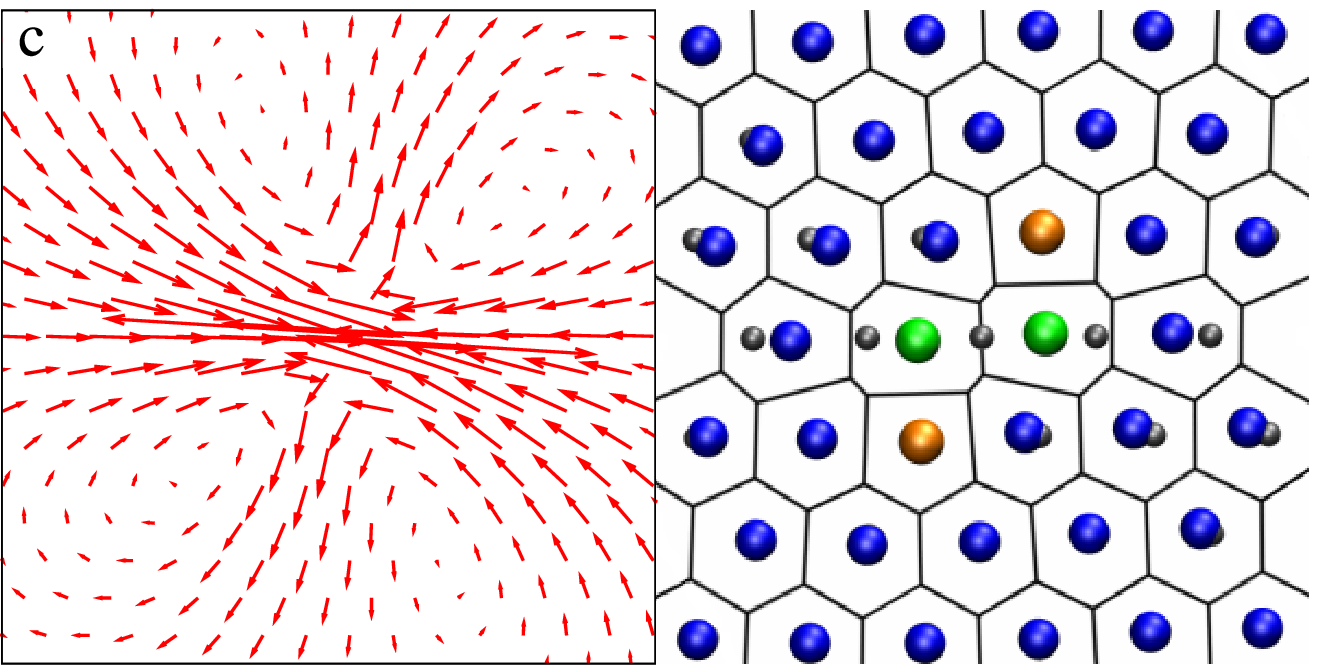}}
\caption{Displacement fields (left hand side) and particle configurations (right hand side) of the $V_2$ vacancy (a), and the $V_3$ vacancy (b) and the $V_a$ vacancy (c). The arrows representing the particle displacements are exaggerated in length by a factor of 20 for better visibility. The color code is the same as in Fig.\ref{fig:displacementinter}.}
\label{fig:displacementvac}
\end{figure}

Also vacancies can occur in various configurations with displacement fields displaying  quite complex patterns and symmetries lower than that of the underlying lattice. Three minimum energy configurations are shown in  Fig. \ref{fig:displacementvac}. In the the vacancy configuration $V_2$ (see Fig. \ref{fig:displacementvac}a), particles move mainly on the $x$-axis to partially fill the void left by a removed particle. As a result, particles above and below the void site move outward generating a vortex-like displacement field. This two-fold vacancy $V_2$ has the same symmetry as the $I_2$ interstitial, but  their displacement patterns are not simply related by inversion. In particular, the vortex structure observed for the vacancy is absent in the interstitial case. In the configuration $V_3$ with threefold symmetry, particles partially fill the vacancy void by moving in along three axes rather than two. On the other three low-index axes, particles are moved outward in response to the removed particle. A vacancy configuration analogous to the $I_d$ interstitial seems not to be stable even at $T=0$. A configuration prepared in this symmetry ends in an antisymmetric configuration $V_a$ (see Fig.\ref{fig:displacementvac}c). Energetically, configurations $V_2$ and $V_a$ are equal and lower than configuration $V_3$ by more than twice the thermal energy $k_{\rm B}T$ at melting thus exceeding the energy difference of the corresponding interstitial configurations by more than an order of magnitude. This energy difference is less than 10\% of the total defect energy.

\section{Elasticity Theory}
\label{sec:elastic}

Near the defect site non-linearities and discrete lattice effects dominate the displacement pattern as evidenced by the highly anisotropic local structure of vacancies and interstitials. Far away from the defect, however, the perturbation of the 2d-crystal should be described accurately by continuum elasticity theory. In this regime, the response of the system to a point defect should depend on the specific form of the interaction potential only through the particular values of the elastic constants. To verify to which extent elasticity theory is valid for two-dimensional colloidal crystals of soft particles, we first review the basic equations of elasticity theory and then solve them for an idealized singular defect model consisting of a pair of singular forces of equal magnitude and opposite direction \cite{ESHELBY,ESHELBY_ACTA,SCATTERGOOD}.  

Linear elasticity theory is usually formulated in terms of the symmetric strain tensor \cite{LANDAU_LIFSCHITZ}
\begin{equation}
\epsilon_{ij} = \frac{1}{2} \left(\frac{\partial u_i}{\partial r_j} + \frac{\partial u_j}{\partial r_i} \right),
 \end{equation}
where $u_i$ and $r_i$ are the $i$-th component of the displacement and the position, respectively. For small strains, Hook's law applies and the stress $\sigma_{ij}$ is linearly related to the strain $\epsilon_{ij}$,
\begin{equation}
\sigma_{ij}=C_{ijkl}\epsilon_{kl},
\label{equ:Hook}
\end{equation}
where $C_{ijkl}$ is the stiffness tensor. Here and in the following, summation over repeated indices is implied. For isotropic materials, such as two-dimensional crystals with triangular lattice, this relation reduces to
\begin{equation}
\sigma_{ij} = \lambda \delta_{ij}\epsilon_{kk} + 2\mu \epsilon_{ij}
\label{equ:Hook_Isotropic},
\end{equation}
where $\lambda$ and $\mu$ are the so-called Lam\'e coefficients. The Lam\'e coefficient $\mu$ is also called the shear modulus. 

In order to calculate the displacement field generated by a point defect one must be able to determine how the elastic continuum reacts to external forces. The condition that the forces on each infinitesimal volume element balance leads to
\begin{equation}
\frac{\partial \sigma_{ij}}{\partial r_j} +f_i=0,
\label{equ:equil_stress} 
\end{equation}
where $f_i$ is component $i$ of a given volume force $f({\bf r})$ acting at ${\bf r}$. Using the stress-strain relation from Equ. (\ref{equ:Hook_Isotropic}), these equilibrium conditions can be formulated in terms of the strains rather than the stresses, 
\begin{equation}
\lambda \frac{\partial}{\partial r_i}\epsilon_{kk} + 2\mu \frac{\partial \epsilon_{ij}}{\partial r_j}+f_i = 0, 
\label{equ:equil_strain}
\end{equation}
Inserting the definition of the strain into this equation one obtains the equilibrium conditions for the displacement field ${\bf u}({\bf r})$,
\begin{equation}
(\lambda + \mu) \frac{\partial}{\partial r_i} \frac{\partial u_j}{\partial r_j}+\mu \Delta u_i + f_i = 0, 
\label{equ:equil_u}
\end{equation}
where $\Delta = \partial^2 / \partial x^2 + \partial^2 / \partial y^2$ is the Laplace operator. Solving this equation for a particular arrangement of forces used to model the point defect then yields the displacement field caused by the forces.

From the displacement field one can then determine the energetics of the point defect. In terms of the strain tensor and the Lam\'e coefficients the elastic free energy density of the system is given by
\begin{equation}
\label{equ:free_energy_density}
g = \frac{\lambda}{2}\epsilon^2_{kk} + \mu \epsilon_{ij}^2.
\end{equation}
Accordingly, the energy density at $T=0$ is given by 
\begin{equation}
\label{equ:energy_density}
e = \frac{\lambda}{2}\epsilon^2_{kk} + \mu \epsilon_{ij}^2 - p \epsilon_{kk}.
\end{equation}
The last term of this equation stems from the work done against the pressure $p$ by the dilatation $\epsilon_{kk}$. The strain tensor can also be written as the sum of a trace-free shear and a homogeneous dilation leading to the expression  
\begin{equation}
g = \mu \left(\epsilon_{ij}-\frac{1}{2}\delta_{ij}\epsilon_{kk}\right)^2+\frac{K}{2}\epsilon_{kk}^2,
\label{equ:elastic}
\end{equation}
where $K$ is the so-called bulk modulus related to $\lambda$ and $\mu$ by 
\begin{equation}
K=\lambda + \mu.
\end{equation}
The Poisson ratio $\nu$, i.e., the negative ratio of transverse strain to axial strain upon uniaxial loading, is given by
\begin{equation}
\nu = \frac{\lambda}{\lambda + 2\mu}=\frac{K-\mu}{K+\mu}
\end{equation}
and describes how a material reacts when stretched. In the next subsection we will calculate the elastic constants for our system at $T=0$.
 
\subsection{Elastic moduli}

For a crystal in which particles interact with a pair potential $v(r)$ depending only on the interparticle distance $r$ the total energy $E$ of $N$ particles is given by
\begin{equation}
E = \frac{1}{2}\sum_{i \neq j} v(|{\bf r}_i-{\bf r}_j|),
\end{equation}
where ${\bf r}_i$ and ${\bf r}_j$ are the positions of particles $i$ and $j$ respectively. In this case and for $T=0$, the energy density $e_0$ of the undistorted lattice, the pressure $p$, as well as the elastic constants $K$ and $\mu$ can be calculated from simple lattice sums:
\begin{equation}
e_0 = \frac{\rho}{2}\sum_{i}{'}v(r_i),
\end{equation}
\begin{equation}
p = -\frac{\rho}{4}\sum_{i}{'}v'(r_i)r_i,
\end{equation}
\begin{equation}
K = \frac{\rho}{8}\sum_{i}{'}\left[v''(r_i)r_i^2-v'(r_i)r_i\right],
\end{equation}
and 
\begin{equation}
\mu = \frac{\rho}{2}\sum_{i}{'}\left[v''(r_i)\left(\frac{x_iy_i}{r_i}\right)^2+v'(r_i)\frac{y_i^4}{r_i^3}\right].
\end{equation}

Here, $v'(r)$ and $v''(r)$ are the first and second derivative of the pair potential, respectively, $\rho$ is the number density, $r_i$ is the distance of particle $i$ from the origin, and $x_i$ and $y_i$ are its Cartesian coordinates. The lattice position is chosen such that there is one particle at the origin. The sums in the above equations must include a sufficient number of particles to ensure convergence of the sums. (The prime on the sum symbol indicates that the particle at the origin is not included in the sum).  

\begin{figure}[htb,floatfix]
\centerline{\includegraphics[width=7.0cm]{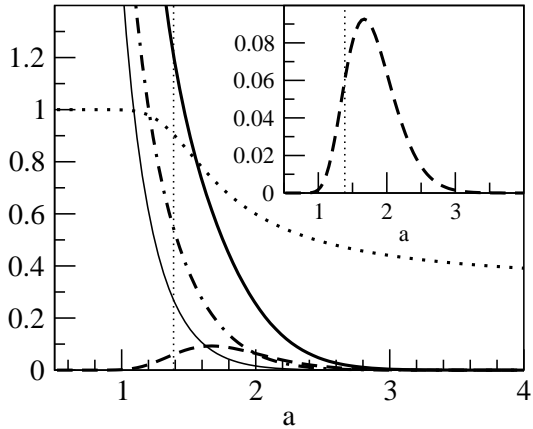}}
\caption{Bulk modulus $K$ (solid line), shear modulus $\mu$ (dashed line), pressure $p$ (dash-dotted line), energy density $e$ (thin solid line), and Poisson ratio $\nu$ (dotted line) as a function of the lattice constant $a$. The moduli are given in units of $\varepsilon / \sigma^2$ and the lattice constant in units of $\sigma$. In the inset, the shear modulus $\mu$ is displayed on a larger scale. The vertical thin dotted line indicates the lattice constant $a=1.3872 \sigma$ corresponding to the density $\rho=0.6 \sigma^{-2}$, at which all calculations discussed in this article are carried out.}
\label{fig:moduli}
\end{figure}

The elastic constants $\mu$, and $K$ as well as the Poisson ratio $\nu$, the pressure $p$ and the energy density $e$, calculated using such sums, are shown in Fig. \ref{fig:moduli} as a function of the lattice constant $a$, which, in a triangular lattice, is related to the density by $\rho=2/{\sqrt{3}a^2}$. At a density of $\rho=0.6\sigma^2$, the density at which all calculations presented in this article are carried out, the elastic constants have the values $K=1.2089\, \varepsilon \sigma^{-2}$ and $\mu=0.060183 \,\varepsilon \sigma^{-2}$, the pressure is $p=0.544245 \, \varepsilon \sigma^{-2}$, and the Poisson ratio is $\nu=0.905151$. The energy density is $e_0=0.269125 \,\varepsilon \sigma^{-2}$ corresponding to an energy per particle of $E/N=0.448542 \varepsilon$. Note that at this density, the system is stabilized against shear only by interactions beyond nearest-neighbor contributions; estimation of $\mu$ from nearest neighbor interactions only yields a negative shear rate at this density. While the bulk modulus $K$ increases monotonically with the density (and, as the pressure $p$, is proportional to $\rho^2$ for small densities), the shear modulus reaches a maximum at $a \approx 1.67$ and then rapidly decays to zero for lattice constants larger and smaller than that. This behavior of the shear modulus is a reflection of the phenomenon of reentrant melting observed in the three-dimensional Gaussian core model \cite{STILLINGER_GCM,GAUSSIAN_CORE_PHASE} and indicates that also in two dimensions a Gaussian core crystal melts if sufficiently compressed.

\subsection{Point defect model}
\label{sec:model}

We next use elasticity theory to determine the displacement field created by introducing an idealized point defect into a perfect isotropic crystal. The dilatation (or contraction) caused by the defect is modeled by two orthogonal pairs of forces. Each pair consists of two forces of equal magnitude $F$ but opposite directions acting at two points separated by the distance $h$. If one assumes that one force pair acts in $x$-direction and the other one in $y$-direction and that the defect is centered a the origin, the total force density is given by 
\begin{eqnarray}
{\bf f}({\bf r})&=&-F\delta({\bf r}){\bf e}_x+F\delta({\bf r}-h{\bf e}_x){\bf e}_x \nonumber \\ 
& & -F\delta({\bf r}){\bf e}_y+F\delta({\bf r}-h{\bf e}_y){\bf e}_y. 
\label{equ:f_model}
\end{eqnarray}
Here, ${\bf e}_x$ and ${\bf e}_y$ are the unit vectors in $x$- and $y$-direction, respectively, and $\delta ({\bf r})$ is the Dirac delta-function in two dimensions. One then lets the separation $h$ go to zero and the force $F$ go to infinity in a way such that $Fh$ remains constant. This defect model, in which the net force acting on the material vanishes, is equivalent to inserting a small circular inclusion into a hole of different size \cite{ESHELBY_ACTA}. 

The displacement field caused by this type of point defect can be determined by first calculating the Green's function for a singular force and than carrying out the limit $h\rightarrow 0$. Alternatively, one can carry out the limit $h\rightarrow 0$ first and then solve the equilibrium condition for the force density obtained in that way. In the following we will calculate the displacement field of the point defect model using this second approach, in which periodic boundary conditions can be taken into account particularly easily.

Carrying out the limit $h \rightarrow 0$ as described above the total force density of Equ. (\ref{equ:f_model}) reduces to
\begin{equation}
{\bf f}({\bf r})=-Fh \nabla \delta ({\bf r}).
\end{equation}
Inserting this expression into Equ. (\ref{equ:equil_u}) one obtains the equilibrium condition for this simple point defect model,
\begin{equation}
(\lambda + \mu) \frac{\partial}{\partial r_i} \frac{\partial u_j}{\partial r_j}+\mu \Delta u_i=Fh \frac{\partial}{\partial r_i} \delta ({\bf r}).
\label{equ:equil_u_point}
\end{equation}
Taking the divergence on both sides yields
\begin{equation}
\Delta (\lambda + 2\mu) \frac{\partial u_j}{\partial r_j}=Fh \Delta \delta ({\bf r}).
\end{equation}
To solve this equation it suffices to find a displacement field that obeys
\begin{equation}
(\lambda + 2\mu) \frac{\partial u_j}{\partial r_j}=Fh \delta ({\bf r}).
\label{equ:after_div}
\end{equation}
Using the Helmholtz-decomposition in two dimensions, we now write the displacement in terms of the gradients of two scalar functions $\phi({\bf r})$ and $A({\bf r})$ as a sum of an irrotational and a divergence-free part, 
\begin{equation}
u_i = \frac{\partial \phi}{\partial r_i} + \omega_{ij} \frac{\partial A}{\partial r_j}, 
\label{equ:Helmholtz}
\end{equation}
where the matrix $\omega_{ij}$ exchanges the components of the gradient and changes the sign of one of them: $\omega_{11}=\omega_{22}=0$ and $-\omega_{21}=\omega_{12}=1$. Then, Equ. (\ref{equ:after_div}) becomes
\begin{equation}
\label{equ:Poisson}
\Delta \phi ({\bf r})=2 \pi \gamma \delta ({\bf r}),
\end{equation}
where we have used the fact that $\omega_{ij} \partial A / \partial r_j$ is divergence-free and the parameter $\gamma$, which has the dimension of an area, is given by 
\begin{equation}
\gamma = \frac{Fh}{2\pi (\lambda + 2\mu)}.
\end{equation}
Equation (\ref{equ:Poisson}) is the Poisson equation of electrostatics for a point charge of strength $-\gamma$ in two dimensions. 

A similar equation can be derived for the the scalar function $A({\bf r})$ by taking the 2d-vorticity, defined as $\omega_{ij}\partial v_j / \partial r_i$ for an arbitrary vector field ${\bf v}=(v_1, v_2)$, of both sides of Equ. (\ref{equ:equil_u_point}). Since the vorticity of a gradient field vanishes, one obtains the biharmonic equation 
\begin{equation}
\mu \Delta (\Delta A({\bf r}))=0.
\end{equation}
This equation is obeyed if the scalar field $A({\bf r})$ is a solution of the Laplace  equation 
\begin{equation}
\Delta A({\bf r})=0.
\end{equation}
In the following, we will use the trivial solution $A({\bf r})={\rm const}$ and satisfy the boundary conditions through proper solution of the Poisson equation (\ref{equ:Poisson}) for the  scalar field $\phi$.

To do that, we note that $K({\rm r})=\ln (r)/2 \pi$ is a solution of $\Delta K = \delta ({\bf r})$ (see, for instance, Ref. \cite{COURANT}), and hence we obtain the Green's function
\begin{equation}
\label{equ:phi_logarithmic}
\phi(r)=\gamma \ln (r).
\end{equation}
The corresponding displacement field ${\bf u}({\bf r})$ follows by differentiation according to Equ. (\ref{equ:phi_logarithmic}),
\begin{equation}
\label{equ:u_oneoverr}
{\bf u}({\bf r}) = \gamma \frac{\bf r}{r^2}.
\end{equation}
Thus, the displacement field caused by the point defect is isotropic and long-range with a  magnitude that is proportional to $1/r$. This result is valid for an infinitely extended elastic medium where the boundary conditions ${\bf u}=0$ apply at infinity. This situation, however, does not correspond to the boundary conditions applied in computer simulations. In the following section we will discuss how to solve Equ. (\ref{equ:Poisson}) with the appropriate boundary conditions.  

\subsection{Boundary conditions}

In comparing the results of particle simulations with those of elasticity theory it is important to realize that the displacement fields predicted by continuum theory are of long-range nature. Therefore, it is crucial that corresponding boundary conditions are used in both cases. All simulations discussed in this paper are done with periodic boundary conditions in order to minimize finite size effects and preserve the translational invariance of the perfect lattice. Hence, also the continuum calculations need to be carried out with periodic boundary conditions. For a rectangular system with side lengths $L_x$ and $L_y$, periodic boundary conditions require that ${\bf u}({\bf r})={\bf u}({\bf r}+{\bf l})$, where ${\bf l}=(iL_x, jL_y)$ is an arbitrary lattice vector with integer $i$ and $j$. In the following, we will solve the Poisson equation (\ref{equ:Poisson}) for this type of boundary conditions. 

We start by noting that the homogenous part of the Poisson equation (\ref{equ:Poisson}) admits the non-trivial solution $\phi_0({\bf r})={\rm const}$ that satisfies the boundary conditions. Therefore, one needs to consider the {\em extended} Green's function for the solution of the general Poisson equation $\Delta \phi({\bf r})=2\pi \rho({\bf r})$ \cite{COURANT,NEUMANN_PC}. In this case, the right hand side of the Poisson equation must be orthogonal to the solution $\phi_0({\bf r})$,
\begin{equation}
\label{equ:condition_1}
\int {\rm d}{\bf r}\,\phi_0({\bf r})\rho({\bf r})={\rm const} \int {\rm d}{\bf r}\,\rho({\bf r})=0.
\end{equation}
In electrostatics, this condition corresponds to charge neutrality (the physical meaning of this condition in our case will be discussed below). To satisfy this orthogonality condition we must modify the Poisson equation (\ref{equ:Poisson}) by subtracting $1/A$ from the delta function,
\begin{equation}
\label{equ:extended_Poisson}
\Delta \phi ({\bf r})=2 \pi \gamma \left[\delta ({\bf r})-\frac{1}{A}\right],
\end{equation}
where $A$ is the area of the rectangular basic cell. In this modified equation, the right hand side contains a homogeneous ``neutralizing background'' that exactly compensates for the ``charge'' of the delta function. Solution of this equation yields the extended Green's function of the problem. To obtain a unique solution  $\phi({\rm r})$ of this equation one must furthermore require that this solution be orthogonal to $\phi_0({\rm r})$,
\begin{equation}
\label{equ:normal}
\int {\rm d}{\bf r}\,\phi_0({\bf r})\phi({\bf r})={\rm const} \int {\rm d}{\bf r}\,\phi({\bf r})=0.
\end{equation}
For our case this condition is irrelevant, as only derivatives of $\phi({\bf r})$ carry physical significance. Once the function $\phi({\rm r})$ has been determined by solving Equ. (\ref{equ:extended_Poisson}), the displacement field follows by differentiation.

\subsubsection{Rigid circular container}
\label{sec:rigid}

Before we embark on the solution of the extended Poisson equation (\ref{equ:extended_Poisson}) for periodic boundary conditions, we illustrate the concepts introduced above by determining the displacement field of a point defect in an elastic material enclosed in a container with rigid walls. Due to these walls, the component of the displacement field normal to walls must vanish at the wall, $u_\perp = 0$. No condition applies for the parallel component $u_{||}$. For a rectangular container such rigid wall boundary conditions are equivalent to periodic boundary conditions. If we assume, without loss of generality, that the point defect is located at the center of the rectangular periodic cell, the component of the displacement field normal to the boundary of the periodic cell vanishes also in this case. In the following, we will determine the effect of such rigid boundary conditions on the displacement field of a point defect located at the center of a circular cavity enclosed by hard walls. For this case, which exhibits all complications mentioned above, a simple analytical solution can be easily obtained.   

Consider a two-dimensional elastic isotropic material enclosed in a circular container of radius $R$. We choose the coordinate system such that the origin is at the center of the container. To determine the displacement field caused by a point defect of strength $\gamma$ placed at the origin we need to solve Equ. (\ref{equ:extended_Poisson}) under the condition that the at distance $R$ from the origin the displacement $u_\perp$ normal to the wall vanishes. We construct a solution by superposing the solution for the extended material, ${\bf u}_\infty({\bf r})=\gamma {\bf r}/r^2$, and a homogeneous contraction, ${\bf u}_{\rm c}({\bf r})=-\alpha {\bf r}$,
\begin{equation}
{\bf u}({\bf r})={\bf u}_\infty({\bf r})+{\bf u}_{\rm c}({\bf r})=\gamma \frac{{\bf r}}{r^2}-\alpha {\bf r}.
\end{equation}
While ${\bf u}_{\rm c}({\bf r})$ corresponds to a homogeneous contraction without shear, ${\bf u}_\infty({\bf r})$ corresponds to a pure shear without dilatation (except for $r= 0$). To satisfy the boundary conditions at $r=R$, we set $\alpha=\gamma / R^2$ obtaining
\begin{equation}
\label{equ:displacement_circular}
{\bf u}({\bf r})=\gamma {\bf r}\left(\frac{1}{r^2}- \frac{1}{R^2}\right).
\end{equation}
This displacement field corresponds to the ``potential'' 
\begin{equation}
\label{equ:potential_sum}
\phi({\bf r})=\phi_\infty({\bf r})+\phi_{\rm c}({\bf r})=\gamma \left(\ln \frac{r}{R}-\frac{1}{2}\frac{r^2}{R^2}+\frac{3}{4} \right),
\end{equation}
where the last constant on the right hand side takes care of the condition expressed in Equ. (\ref{equ:normal}). It is straightforward to verify that 
\begin{equation}
\Delta \phi ({\bf r})=\Delta \phi_\infty ({\bf r})+\Delta \phi_c ({\bf r})=2\pi \gamma \delta ({\bf r})-\frac{2\pi \gamma}{A},
\end{equation}
such that the potential from Equ. (\ref{equ:potential_sum}) satisfies the extended Poisson equation (\ref{equ:extended_Poisson}). 

\begin{figure}[htb,floatfix]
\centerline{\includegraphics[width=7.0cm]{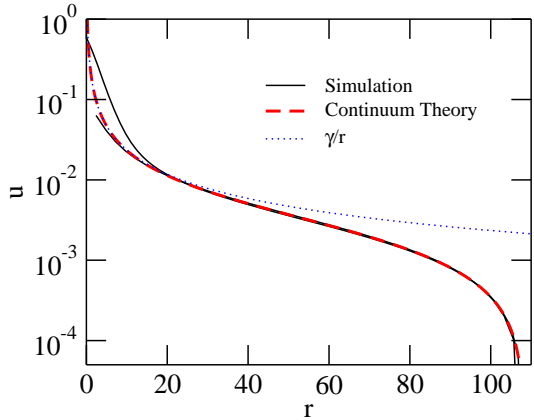}}
\caption{Displacement components $u_x$ and $u_y$  as a function of the distance $r$ (thin solid lines) for an $I_2$ interstitial with its main axis oriented in $x$-direction. The circular hard wall container has a radius of $R=106.8 \sigma$. Also plotted is the displacement computed from continuum theory according to Equ. (\ref{equ:displacement_circular}) (dashed line) for a defect strength of $\gamma=0.234495 \sigma^2$ which best fits the numerical results in the far field and the simple $1/r$-behavior (dotted line). }
\label{fig:absdisplacement_circular}
\end{figure}

\begin{figure}[htb,floatfix]
\centerline{\includegraphics[width=8.0cm]{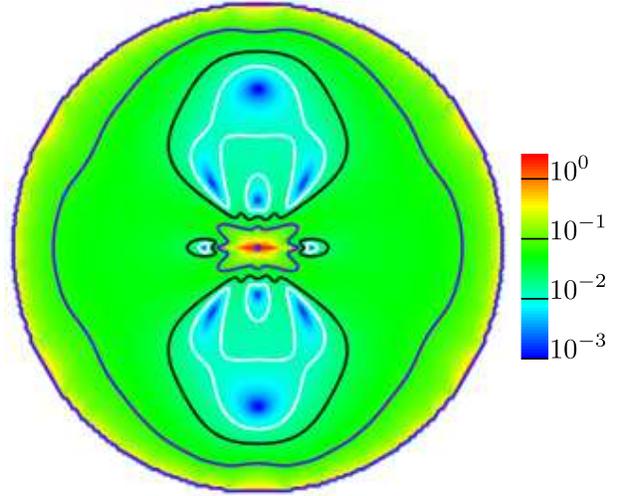}}
\caption{Color coded relative deviation $\xi$ calculated according to Equ. (\ref{equ:relative_error}) of an $I_2$ interstitial at the origin of a rigid circular box with radius $R=108.19\sigma$. The fit parameter $\gamma = 0.234495\sigma^2$ was found by minimizing the sum of the relative error of particles at distances larger than $30\sigma$. The contour lines in white, black, and blue represent an error of $1\%$, $2\%$, and $5\%$, respectively.}
\label{fig:reldevround2}
\end{figure}

A comparison of results obtained numerically for an $I_2$ interstitial and the prediction of elasticity theory (Equ. (\ref{equ:displacement_circular})) is shown in Fig. \ref{fig:absdisplacement_circular}. For the particle system the rigid container was realized by carrying out the calculation in a larger system in which all particles beyond a distance of $R$ from the origin where kept at fixed positions. The displacements obtained from particle and continuum calculations agree very well for all defect distances larger than about 15 lattice constants. In Fig. \ref{fig:reldevround2} we depict the relative error which we define in the following way:
\begin{equation}
\xi \equiv \frac{|{\bf u}_p({\bf r})-{\bf u}_c({\bf r})|}{|{\bf u}_c({\bf r})|}.
\label{equ:relative_error}
\end{equation}
Here, ${\bf u}_p({\bf r})$ and ${\bf u}_c({\bf r})$ are the displacement fields obtained from the particle system and from the predictions of continuum theory, respectively. The relative error close to the defect can be larger than $100\%$ (red). For distances of $r>20\sigma$ with find an error of approximately $1-5\%$ (green). Close to the rigid container the relative error increases again due to discrete lattice effects. 

\begin{figure}[htb,floatfix]
\centerline{\includegraphics[width=7.0cm]{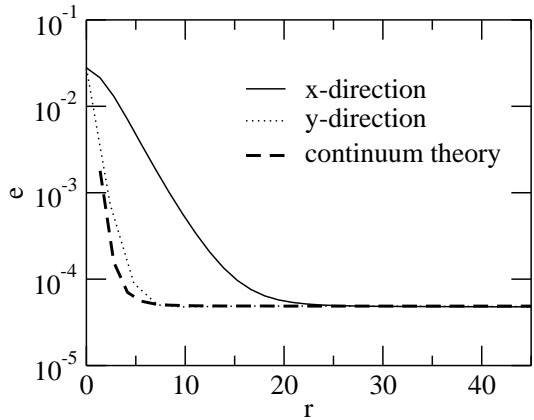}}
\caption{Energy density $e$ as a function of distance $r$ measured along the $x$-axis (solid line) and the $y$-axis (dotted line) for the particle system in a circular container with hard walls and radius $R=71.91\sigma$. Also shown is the energy density calculated from continuum theory according to Equ. (\ref{equ:density_circular}) (dashed line).}
\label{fig:density_circular}
\end{figure}

Similar agreement is found also for the energy density as shown in Fig. \ref{fig:density_circular}. For the displacement field of Equ. (\ref{equ:displacement_circular}) one finds, using Equ. (\ref{equ:energy_density}) the energy density
\begin{equation}
e(r) = 2\mu \frac{\gamma^2}{r^4}+2K \frac{\gamma^2}{R^4}+2p\frac{\gamma}{R^2}.
\label{equ:density_circular}
\end{equation}
This prediction of elasticity theory matches the energy density determined numerically in $y$-direction (see Fig. \ref{fig:density_circular}). Due to the strong anisotropy of the $I_2$ defect, larger deviations are observed in $x$-direction. For distances of more than about 15 lattice spacings the energy density plateaus at a constant value. In this regime, the energy density is essentially constant, $e=2p\gamma/R^2$, and corresponds to the work done by the defect against the pressure $p$. As discussed below, the plateau value of the density is related to the neutralizing background on the right hand side of Equ. (\ref{equ:extended_Poisson}).

\subsubsection{Ewald summation}

For an isolated defect at the center of a rectangular cell, the symmetry imposed by periodic boundary conditions requires that the components of the displacement field orthogonal to the surface of the cell vanish. In a first attempt to obtain the displacement field for such boundary conditions one may start from the solution for the extended material and satisfy the boundary conditions by placing ``image defects'', each of which carries the displacement field for the infinitely extended material, at appropriate positions. For a rectangular cell, an infinite number of image defects arranged on a regular lattice with lattice constants $L_x$ and $L_y$ in $x$- and $y$-direction, respectively, are required. These image defects, which are analogous to the image charges of electrostatics, correspond to the defects in the periodic images of the basic simulation cell. Superposition of the displacement fields of all image defects then yields the displacement field for periodic boundary conditions. 
      
Due to the long-range nature of the defect field for the infinite material, however, such a  summation of the contribution of all image defects leads to displacement fields that are only conditionally convergent. A more appropriate treatment that avoids this problem consists in determining the Green's function of the Poisson equation (\ref{equ:extended_Poisson}) for periodic boundary conditions. The requirement imposed by the periodic boundary conditions can be easily satisfied by expressing the solution as a Fourier series and solving the Poisson equation in Fourier space. This treatment, however, leads to series that are only conditionally convergent with values that depend on the summation order. The solution of this problem using so called Ewald sums is known from electrostatics \cite{LEEUW,NEUMANN,NEUMANN_PC} and consists in separating the conditionally convergent series into a real space and and a Fourier space part,
\begin{eqnarray}
	\label{equ:EwaldPotential}
&&	\hspace{-1.2cm}\phi({\bf r})  = \gamma\left\{\frac{1}{2}\sum_{\bf l}E_i[-\eta^2 |{\bf r}+{\bf l}|^2] \right. 
	\nonumber\\
&&	 	\hspace{0.0cm} \left. - \frac{2 \pi}{A} \sum_{{\bf k} \neq 0} \frac{e^{-k^2/4\eta^2}}{k^2} \cos({\bf k}\cdot {\bf r})  +\frac{\pi}{2 \eta^2 A} \right\}.
\end{eqnarray}
Here,  $E_i(x)=\int_{-\infty}^x (e^t /t) \, dt$ is the exponential integral and $A$ is the area of the rectangular cell. The first sum is over all lattice vectors ${\bf l}$ and the second sum is over all reciprocal vectors ${\bf k}$ consistent with the periodic boundary conditions. The adjustable parameter $\eta$, set to a value of $\eta=6/L_x$ here, determines the rate of convergence of the two sums, but the value of the sums does not depend on $\eta$. The exclusion of the ${\bf k}=0$ term in the above equation stems from the requirement that both $\phi({\bf r})$ and the right hand side of the Poisson equation need to be orthogonal to $\phi_0({\bf r})$ as expressed in Eqs. (\ref{equ:condition_1}) and (\ref{equ:normal}). It is easy to show by direct calculation of the Laplacian $\Delta \phi$ that the above expression for $\phi({\bf r})$ indeed obeys Equ. (\ref{equ:extended_Poisson}) and thus implies a neutralizing background of magnitude $\gamma/A$ as in the previous example. 

From Equ. (\ref{equ:EwaldPotential}) for the scalar function $\phi({\bf r})$ the displacement field of a point defect in a system with periodic boundary conditions is determined by differentiation,
\begin{eqnarray}
	\label{equ:EwaldDisplacement}
&& \hspace{-1.2cm}	u_i({\bf r}) = \gamma \left\{
	\sum_{\bf l} \frac{r_i+l_i}{|{\bf r}+{\bf l}|^2} e^{-\eta^2 |{\bf r}+{\bf l}|^2} \right. \nonumber \\
 &  &	\hspace{0.0cm}+\left. 
	\frac{2\pi}{A} \sum_{{\bf k} \neq 0} \frac{e^{-k^2/4\eta^2}}{k^2} k_i
 \sin(\mathbf{k}\cdot \mathbf{r})\right\}.
\end{eqnarray}
For the systems and parameters considered in this paper, the real space sum can be truncated after the first term and the Fourier space sum can be evaluated accurately using about 50$\times$50 reciprocal vectors. 

As mentioned above, exactly the same boundary conditions apply if the system is constrained by hard walls to reside in an area of given size. Also in that case, the boundary conditions require that the component of the displacement field orthogonal to the walls vanishes. Hence, in the continuum description, hard walls have the same effect as an infinite array of image ``charges'' (plus ``neutralizing background'') placed on a regular lattice with a geometry determined by the wall positions. The effect of such image charges and the neutralizing background is, therefore, not a pure artifact of the periodic boundary condition applied in the simulations, but occurs also in experimental realizations of colloidal crystals of purely repulsive particles which need to be kept together by confining walls. Accordingly, the analysis of displacement patterns (and defect interactions) observed experimentally requires a similar treatment as that used here for the interpretation of our simulation results.    

\subsection{Electrostatic analogy}

In electrostatics, the technique of Ewald summation is used to determine the energetics of periodic systems containing point charges and dipoles. When one uses this technique, one implicitly stipulates that the charges are immersed in a homogeneous background that compensates for the point charges and establishes overall charge neutrality. This neutralizing background is imposed by the periodic boundary conditions; without it, no periodic solution of the Poisson equation exists. Since mathematically the situation we face when determining the displacement field of point defects is identical to that of electrostatics, one may wonder about the physical meaning of the neutralizing background in our Equ. (\ref{equ:extended_Poisson}).

To address this question, we note that the local volume change, or dilatation, due to a displacement field ${\bf u}({\bf r})$ is given be the trace $\epsilon_{kk}$ of the corresponding strain tensor \cite{LANDAU_LIFSCHITZ}. The total change in volume $\Delta V$ of a certain region $G$ is then given as the integral over the dilatation,
\begin{equation}
\Delta V = \int_G {\rm d}{\bf r}\,\epsilon_{kk}({\bf r}).
\end{equation}
On the other hand, it follows from the definition of $\phi$ (see Equ. (\ref{equ:Helmholtz})) that the trace of the strain tensor is equal to the Laplacian of $\phi$,
\begin{equation}
\Delta \phi ({\bf r})=\epsilon_{kk} ({\bf r}).
\end{equation}
Thus, the Poisson equation (\ref{equ:extended_Poisson}) is an equation for the local dilatation. According to this equation, at the defect site the dilatation is required to have a delta like peak of strength $2\pi \gamma$. The total volume change caused by this singular dilatation, $\Delta V=\int_A  {\rm d}{\bf r}\, 2\pi \gamma \delta({\bf r})=2\pi \gamma$, is exactly compensated by the total volume change originating from the constant neutralizing background, $\Delta V=-\int_A  {\rm d}{\bf r}\, 2\pi \gamma / A=-2\pi \gamma$. (Calculating the strain tensor directly from the displacement field of Equ. (\ref{equ:EwaldDisplacement}) indeed yields $\epsilon_{kk}=-2\pi \gamma /A$.) Therefore, the condition of charge neutrality of electrostatics corresponds to the requirement of constant volume in our case. In this analogy, the charge density of electrostatics corresponds to the local dilatation (the charge corresponds to the volume change) and the role of the electric field is played here by the displacement field. 

This interpretation of the Poisson equation (\ref{equ:extended_Poisson}) also suggests a definition of the defect volume $V_d$. As mentioned above, introduction of a point defect of strength $\gamma$ leads to a total volume change which can be viewed as the volume of the defect, 
\begin{equation}
V_d = 2\pi \gamma.
\end{equation}
With periodic (or rigid) boundary conditions the system as a whole is prevented from changing volume and the volume change due to the defect is exactly compensated by the homogeneous neutralizing background. This defect volume is also what one gets when calculating the expansion of a circle under the $1/r$-deformation caused by the idealized defect model (see Equ. (\ref{equ:u_oneoverr})). Measuring the parameter $\gamma$, for instance by fitting the displacement field far from the defect to the continuum theory results,  thus permits to determine the defect volume. For the $I_2$ at a density of $\rho=0.6 \sigma^{-2}$, for example, we found a volume of $V_d=1.44 \sigma^2$, which is slightly smaller then $V_0=1.66 \sigma^2$, the volume per particle in the perfect lattice.

The neutralizing background appearing in the Poisson equation (\ref{equ:extended_Poisson}) also figures in the energy density and can contribute considerably to the total defect energy. According to Equ. (\ref{equ:energy_density}), the energy density includes the term $e_p=-\epsilon_{kk}p$ arising from the work carried out by the defect against the pressure $p$. Away from the singularity at the origin, this component of the energy density is constant, $e_p=2\pi\gamma /A$, and it dominates for large distances from the defect. Although $e_p$ is proportional to $1/A$ and therefore small in general, integrating it over the entire area (leaving out the unphysical singularity at the origin) yields an energy contribution of $E_p=  2\pi\gamma$ which is independent of system size and can be substantial. For an $I_2$ interstitial in our Gaussian core system at $\rho=0.6\sigma^{-2}$, for instance, this contribution amounts to more than 50\% of the total defect energy. Interestingly, no such pressure contribution arises for the displacement field of Equ. (\ref{equ:u_oneoverr}) obtained for the infinitely extended material. Thus, the condition of fixed volume imposed by the periodic boundary conditions leads to measurable effects also in the large system limit and even boundary conditions applied at infinity matter. 


\section{Comparison of simulation and continuum theory}
\label{sec:compare}

In this section we compare the results of the particle-based simulations with the predictions of continuum elasticity theory obtained in the previous section. In particular, we verify at which distances from the point defect elasticity theory becomes valid and which effect boundary conditions have on the displacement fields. We first consider the displacement fields of interstitials, then those of vacancies. 

\subsection{Interstitials}
\label{sec:inter_sim}

\begin{figure}[htb,floatfix]
\centerline{\includegraphics[width=7.0cm]{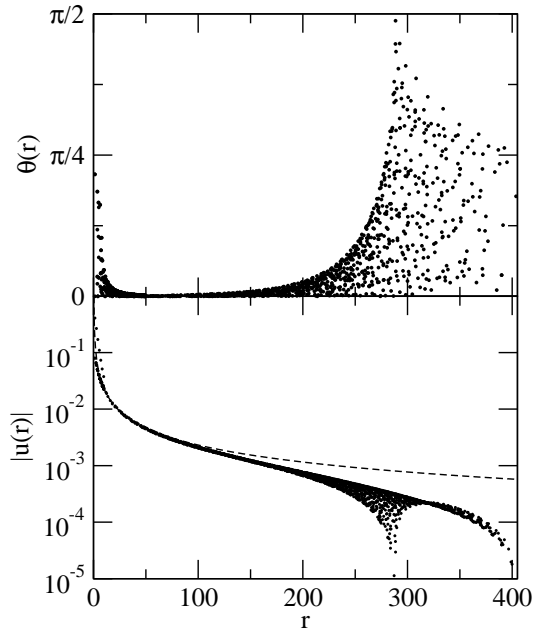}}
\caption{Top: Angle $\theta$ between the displacement vector ${\bf u}$ and the position vector ${\bf r}$ as a function of the distance $r$ from the defect  for an $I_2$ interstitial in a  208$\times$240 particle system. Each dot corresponds to one particle. Bottom: Displacement magnitude $|{\bf u}({\bf r})|$ as a function of $r$. Also shown as a dashed line is the $\gamma/r$-line for $\gamma = 0.23165 \sigma^2$.}
\label{fig:disp_theta}
\end{figure}

As discussed in Sec. \ref{sec:disp}, insertion of an interstitial particle into a perfect lattice can lead to different displacement patterns, all of which are highly anisotropic near the defect site. Farther away from the defect the anisotropy should subside as the isotropic behavior expected from elasticity theory sets in. This is indeed what is observed for intermediate distances from the defect as shown in Fig. \ref{fig:disp_theta} for an $I_2$ defect. In the bottom panel of this figure, the displacement magnitude $|{\bf u}({\bf r})|$ is plotted as a function of the distance $r$ from the defect. Each dot corresponds to one particular particle. While for short distances the displacement magnitude is not a unique function of $r$ due to the anisotropy of the defect, at larger distances  $|{\bf u}({\bf r})|$ is essentially determined by $r$. In this intermediate regime, the displacement magnitude seems to follow the $1/r$-behavior predicted by elasticity theory for the infinitely extended material. Due to the periodic boundary conditions, however, the displacement magnitude cannot remain isotropic as the boundary is approached. In fact, the periodic boundary conditions lead to a spread of $|{\bf u}({\bf r})|$ at larger distances. The two prongs observed in the bottom panel of Fig. \ref{fig:disp_theta} correspond to the directions along the $x$- and $y$-axes and along the diagonals. 

This kind of behavior is observed even more clearly for the displacement directions. In the top panel of Fig. \ref{fig:disp_theta}, the angle $\theta$ between the displacement ${\bf u}({\bf r})$ and the position vector ${\bf r}$ is plotted as a function of distance $r$. As in the bottom panel, each dot corresponds to an individual particle. For an isotropic displacement field, the displacement and the position vector are perfectly aligned and $\theta =0$. Thus, non-zero angles $\theta$ are an indication of anisotropy. For small distances $r$ angles $\theta$ larger than $\pi/4$ occur. For intermediate distances, $20 < r < 150$, the angle $\theta$ is small since in this regime the displacements approximately points away from the origin. For larger distances, the periodic boundary conditions then lead to a spread in $\theta$ and deviations of up to $\theta=\pi /2$ are possible.

\begin{figure}[htb,floatfix]
\centerline{\includegraphics[width=7.0cm]{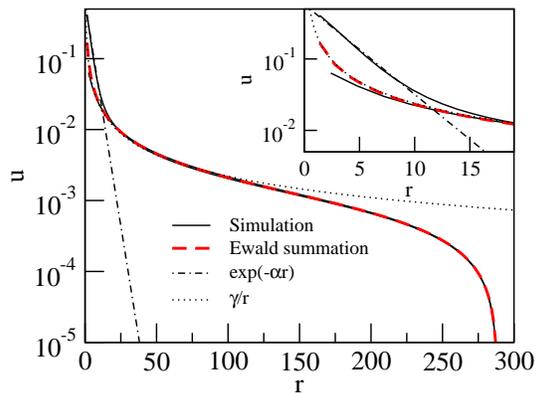}}
\caption{Displacement components $u_x$ and $u_y$ of the $I_{2}$ interstitial as a function of distance $r$ along the $x$-axis and $y$-axis, respectively (solid lines). The main axis of the $I_{2}$ defect is oriented in $x$-direction. Also plotted are the displacement computed from continuum theory by Ewald summation according to Equ. (\ref{equ:EwaldDisplacement}) (dashed line), and simple $1/r$-behavior (dotted line). For short distances, the behavior of the displacement along the $x$-axis is exponential (dash-dotted line) as described by a simple bead-spring model \cite{CODEF}. The inset shows the region close to the defect location. A defect strength of $\gamma = 0.23165 \sigma^2$ was used here since this value yields the best fit of the results obtained from elasticity theory and the numerical results at large distances from the defect.}
\label{fig:absdisplacement}
\end{figure}

\begin{figure}[htb,floatfix]
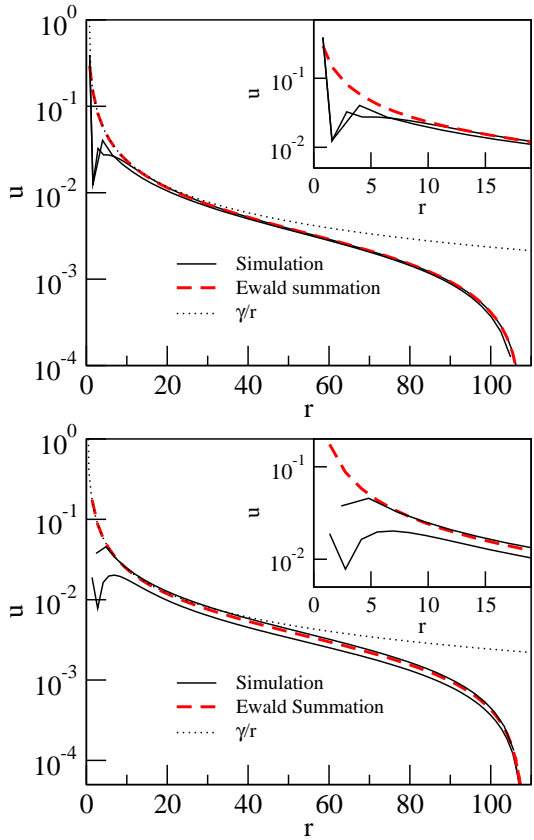

\centerline{\includegraphics[width=7.0cm]{img/displacementcomplete3.eps}}
\centerline{\includegraphics[width=7.0cm]{img/displacementcompleted.eps}}
\caption{Top: Displacement components $u_x$ and $u_y$ of the $I_{3}$ interstitial as a function of distance $r$ along the $x$-axis and $y$-axis, respectively (solid lines). Also plotted are the displacements computed from continuum theory by Ewald summation according to Equ. (\ref{equ:EwaldDisplacement}) (dashed line), and simple $1/r$-behavior (dotted line). The inset shows the region close to the defect location. A defect strength of $\gamma = 0.2347 \sigma^2$ was used. Bottom: Displacement components as above for an $I_d$ interstitial with $\gamma = 0.2425 \sigma^2$.}
\label{fig:reldev3}
\end{figure}

The long-distance behavior described above is perfectly reproduced by linear elasticity theory. As shown in Fig. \ref{fig:absdisplacement} for $I_2$ interstitials and in Fig. \ref{fig:reldev3} for $I_3$ and $I_d$ interstitials, respectively, the displacement calculated  using Ewald summation according to Equ. (\ref{equ:EwaldDisplacement}) agrees very well with the numerical results for all distances larger than about $10-15$ lattice spacings.  In particular, the deviations from the $1/r$-behavior near the cell boundary are perfectly captured by elasticity theory with periodic boundary conditions. For small distances, on the other hand, the displacement field in the particle system is highly anisotropic with strong deviations between the $x$- and $y$-direction. In this non-linear core region elasticity theory is not applicable and the displacements of the three configurations differ. The exponential short-range dependence of $u_x$ on the distance $r$ for $I_2$ interstitials is, however, captured by a simple bead-spring model discussed in Ref. \cite{CODEF}. In this model, the exponential decay constant can be related to the elastic constants of the material. 

\begin{figure}[htb,floatfix]
\centerline{\includegraphics[width=7.0cm]{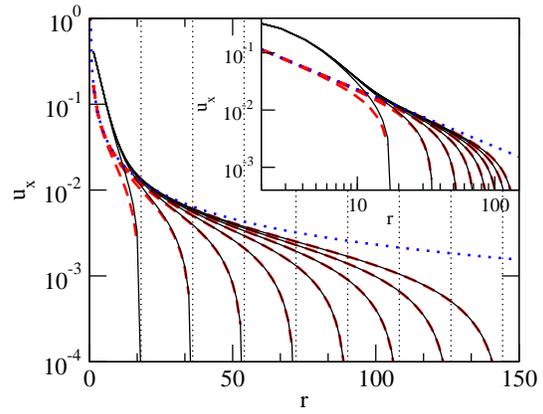}}
\caption{Displacement component $u_x$ as a function of distance $r$ from the defect along the $x$-axis obtained from simulations (solid lines) and according to Equ. (\ref{equ:EwaldDisplacement}) (dashed lines) for system sizes $N = 26\times30, 52\times60, 78\times90, 104\times120, 130\times150, 156\times180,182\times210, 208\times240$. Also shown is the $\gamma / r$ behavior expected in an infinitely extended material (dotted line). The same value of $\gamma=0.23165 \sigma^2$ was used in all cases. The vertical dotted lines indicate the distances of the cell boundaries from the origin for the various system sizes. In the inset the same curves are displayed on a logarithmic scale.}
\label{fig:differentsizes}
\end{figure}

In the comparison of the results obtained for the particle system with those of continuum theory the defect strength $\gamma$ is treated as an adjustable parameter. For each configuration, $I_2$, $I_3$, and $I_d$, the particular displacement strength $\gamma$ was found by optimizing the relative error (see Equ. (\ref{equ:relative_error})) at distances larger than $30.0\sigma$ from the origin of the defect. For the $I_2$ defect, a value of $\gamma=0.23165\sigma^2$ yields the best fit. $I_3$ and $I_d$ interstitials produce a slightly larger displacement with a strength of $\gamma=0.2347\sigma^2$ and $\gamma=0.2425\sigma^2$, respectively. The question arises if this fit is independent on the size of the box. In Fig. \ref{fig:differentsizes} we depict the displacement of an $I_2$ defect for different box sizes together with the results from the Ewald summation. The displacements plotted in the inset of Fig. \ref{fig:differentsizes} on a doubly-logarithmic scale clearly indicate that the algebraic $1/r$ behavior is observed, if at all, only for large system sizes and in a limited distance range.

\begin{figure}[htb,floatfix]
\centering{
\includegraphics[width=7.0cm]{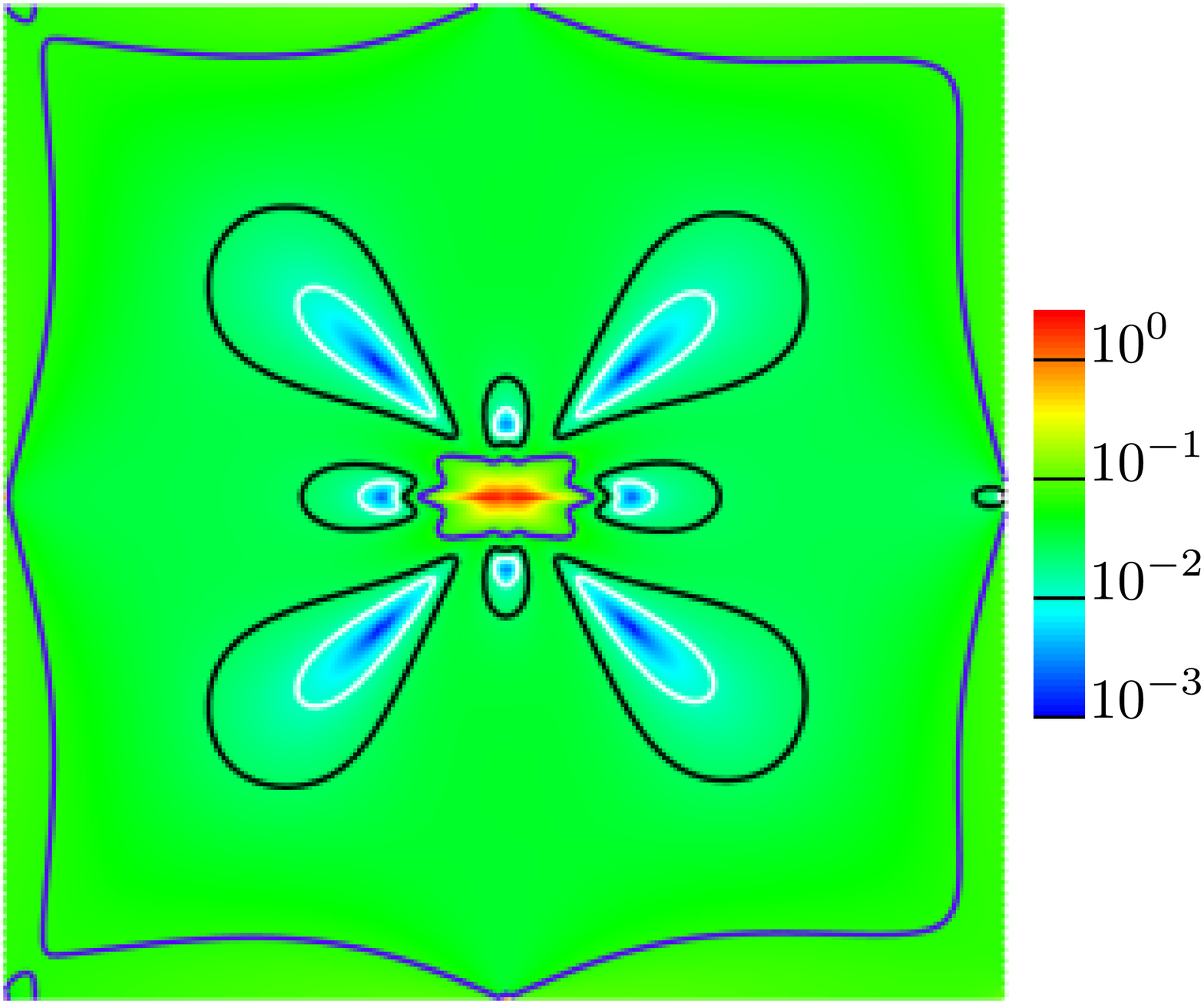}\\
\includegraphics[width=7.0cm]{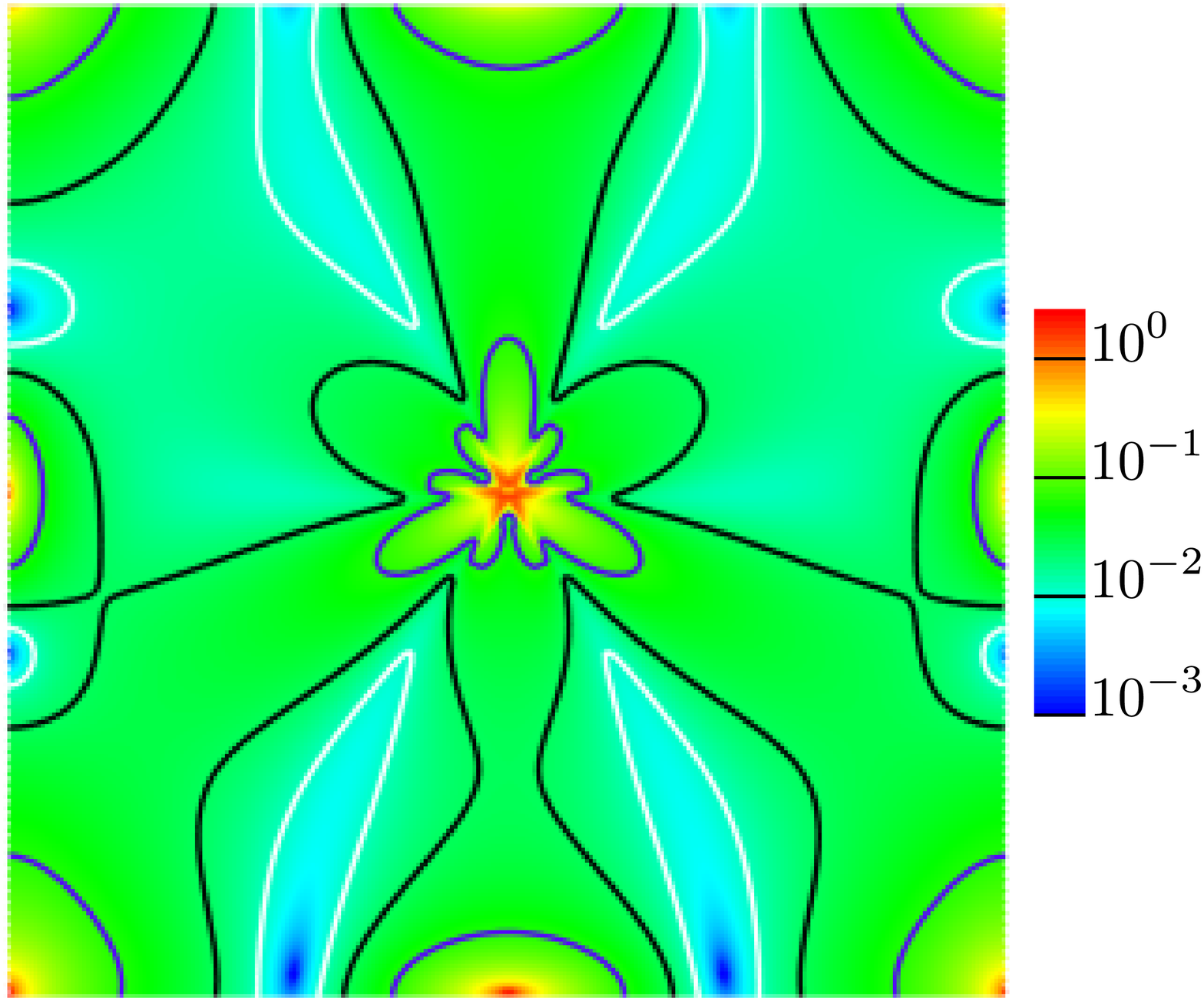}\\
\includegraphics[width=7.0cm]{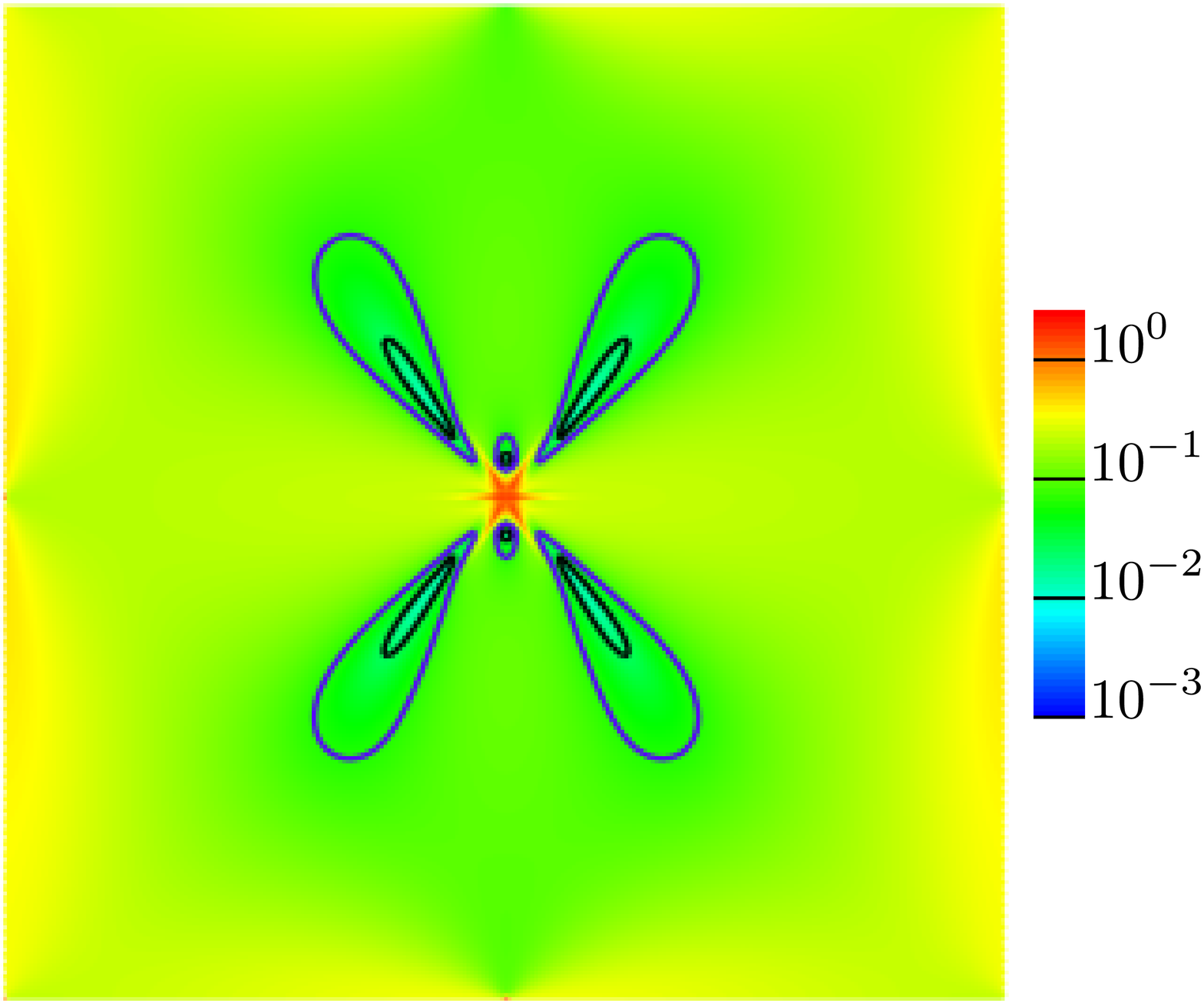}}
\caption{Color coded relative deviations of the displacement ${\bf u}_p({\bf r})$ obtained numerically from the continuum theory prediction ${\bf u}_c({\bf r})$ as a function of $x$ and $y$ calculated according to Equ. (\ref{equ:relative_error}). From top to bottom, the figures depict the relative deviations for the $I_2$ interstitial ($\gamma=0.23165\sigma^2$), the $I_3$ interstitial ($\gamma=0.2347\sigma^2$), and the $I_d$ interstitial ($\gamma=0.2425\sigma^2$). The whole simulation cell of dimensions $L_x=215.28\sigma$ and $L_y=215.12\sigma$ is shown. Colors are assigned on a logarithmic scale which runs from $10^{-3}$ (blue) to $10^{1}$ (red). The white, black, and blue contour lines represent a relative error of $1\%$, $2\%$ and $5\%$, respectively.}
\label{fig:reldev}
\end{figure}

A comparison of the Ewald summation results with the numerical calculations over the whole simulation cell is shown in Fig. \ref{fig:reldev}. The color coded map represents the relative deviation (see Equ. (\ref{equ:relative_error})) of ${\bf u}_p({\bf r})$ from ${\bf u}_c({\bf r})$. In the figure, regions of large and small relative deviation are colored in red and blue, respectively. We find that the $I_2$ and $I_3$ configuration show a relative deviation between $1\%$ to $5\%$ over the whole range. Only in the core region of the defect the deviations are larger. For the $I_d$ defect the deviations are larger and between $10\%$ to $20\%$ also far away from the defect. These deviations are due to discrepancies both in orientations as well as magnitude. 

\begin{figure}[htb,floatfix]
\centerline{\includegraphics[width=7.0cm]{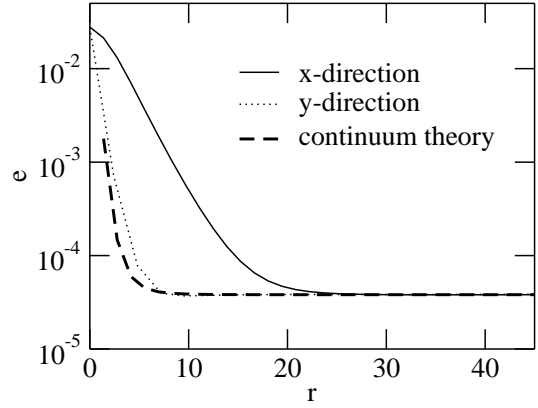}}
\caption{Energy density $e$ of an $I_2$ interstitial in a periodic box as a function of the distance $r$ measured along the $x$-direction (solid line) and the $y$-direction (dotted line). Also shown is the energy density obtained from continuum theory (dashed line) according to Equ. (\ref{equ:energy_density}) for the displacement of Equ. (\ref{equ:EwaldDisplacement}).}
\label{fig:energydensity}
\end{figure}

The energy density of a point defect with periodic boundaries, calculated from Equ. (\ref{equ:energy_density}) for the displacement field of Equ. (\ref{equ:EwaldDisplacement}), is depicted in Fig. \ref{fig:energydensity}. As for the point defect in a circular rigid container discussed in Sec. \ref{sec:rigid}, the energy density becomes constant for large distances. This constant term arises from the work performed by the defect against the external pressure. 

\subsection{Vacancies}

\begin{figure}[htb,floatfix]
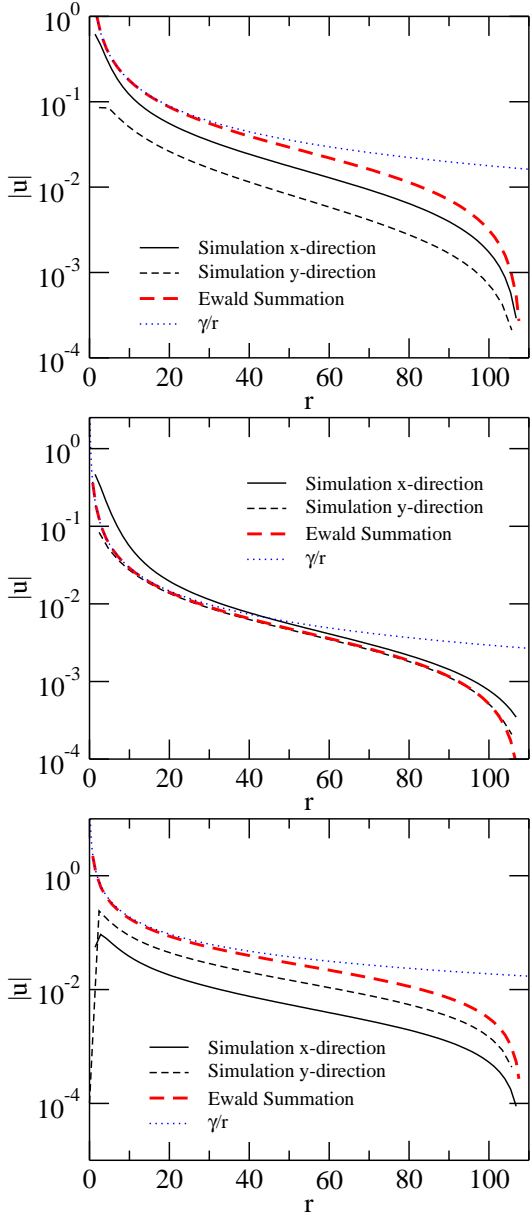

\includegraphics[width=7.0cm]{img/twofoldvac.eps} \\
\includegraphics[width=7.0cm]{img/threefoldvac.eps} \\
\includegraphics[width=7.0cm]{img/antisymetricvac.eps}
\caption{From top to bottom: Absolute value of the displacement components $u_x$ (black solid line) and $u_y$ (black dashed line) of a $V_2$, $V_3$ and $V_a$ vacancy. Also plotted are the absolute values of the predicted displacement field from the Ewald summation Equ. (\ref{equ:EwaldDisplacement}) for the defect strengths $\gamma=-1.78\sigma^2$ ($V_2$), $\gamma=-0.294\sigma^2$ ($V_3$), and $\gamma=-1.89\sigma^2$ ($V_a$), found by minimizing the relative error of Equ. \ref{equ:relative_error} over the whole $xy$-plane (dashed line), as well as $1/r$ behavior (dotted line). }
\label{fig:disp_twofold_vac}
\end{figure}

In the system studied in this paper, vacancies generate displacement patterns that are considerably more intricate than those of interstitials, as can be inferred from a comparison of Figs. \ref{fig:displacementinter} and \ref{fig:displacementvac}. While in the case of interstitials the displacement vectors essentially point away from the defect site, vacancies have displacement fields which point outward or inward depending on the position relative to the defect. For instance, in the  $V_2$ vacancy shown in Fig. \ref{fig:displacementvac}a the displacement vectors point towards the defect site along the $x$-axis, but away from the defect along the $y$-axis. Between the two axes, vortex like structures occur. A similar displacement pattern with alternating displacement directions forms also for the $V_a$ vacancy shown in Fig. \ref{fig:displacementvac}c. This behavior observed in the core region around the defect can not be reproduced by the simple defect model used here for the continuum theory calculations. The displacement field obtained in this model is either oriented towards the defect or away from it depending on the sign of the defect strength $\gamma$. For the $V_3$ vacancy, on the other hand, all displacement vectors point inward and no such complications occur. Nevertheless, the displacement magnitudes shown in Fig. \ref{fig:disp_twofold_vac} for the three vacancy configurations follow qualitatively the form predicted by continuum theory. For all three configurations, varying the defect strength $\gamma$ can lead to better agreement in particular directions (e.g., along the $x$- or $y$-axis), but not over the entire plane. 

\begin{figure}[htb,floatfix]
\centering{
\includegraphics[width=7.0cm]{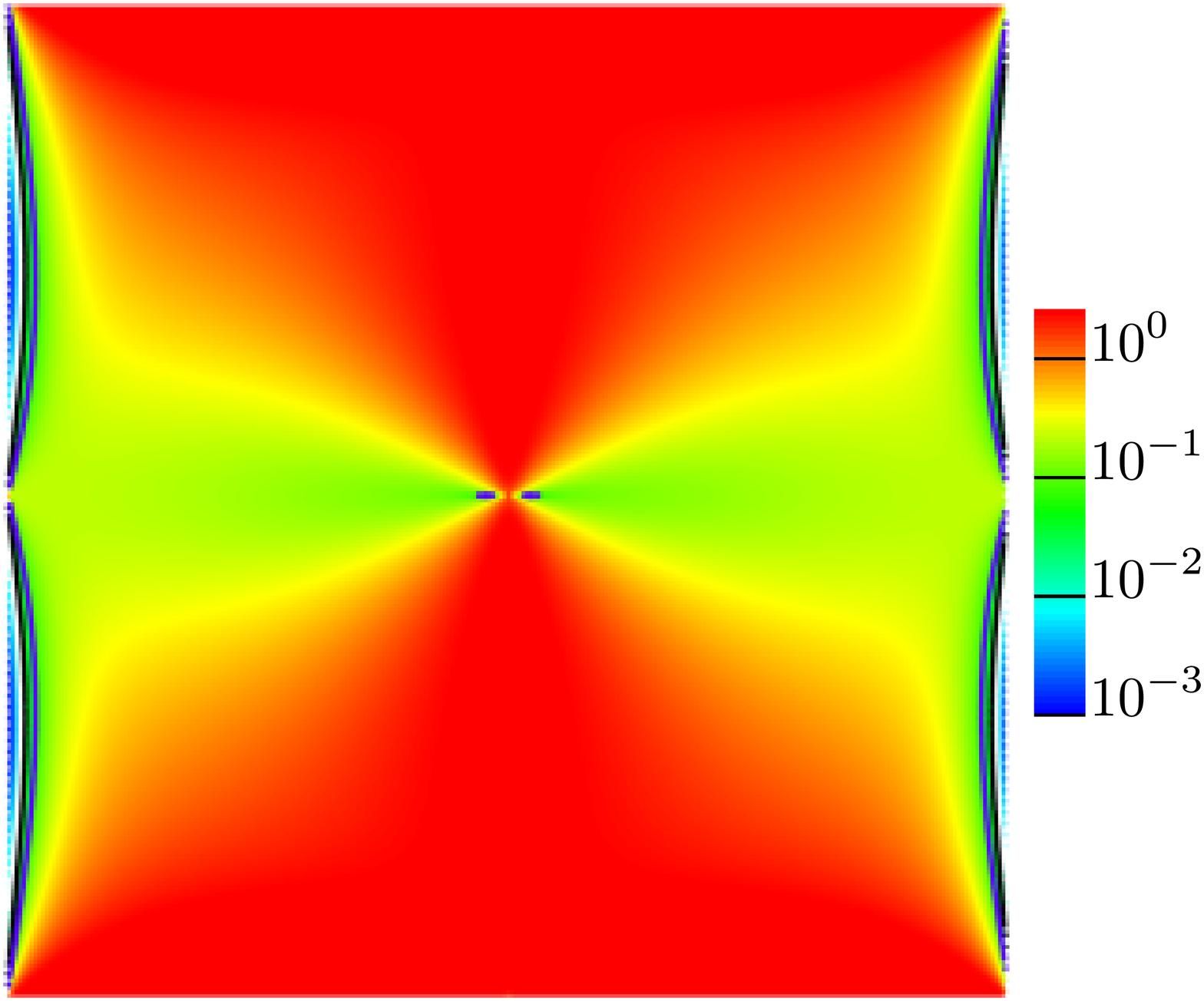}\\
\includegraphics[width=7.0cm]{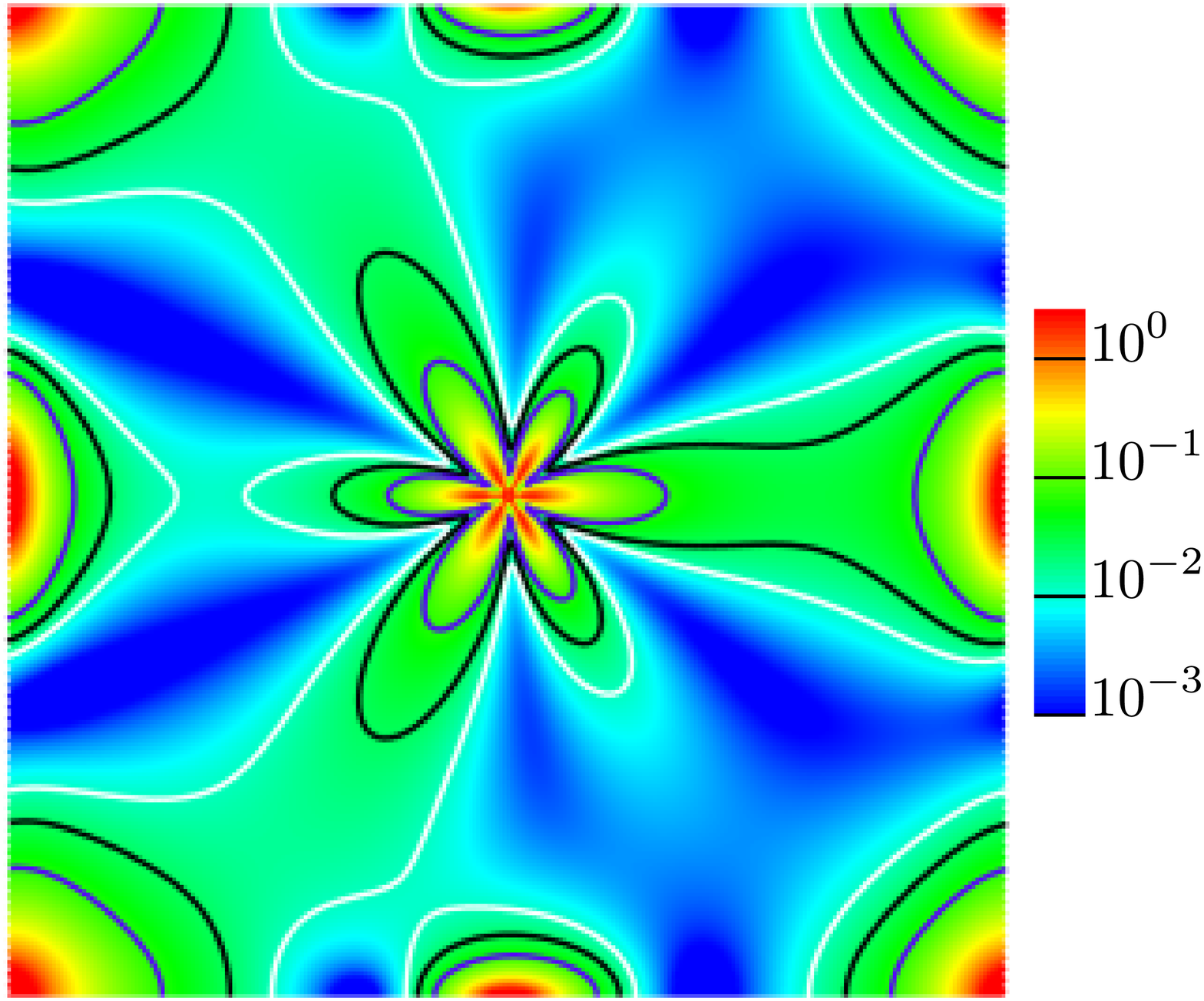}\\
\includegraphics[width=7.0cm]{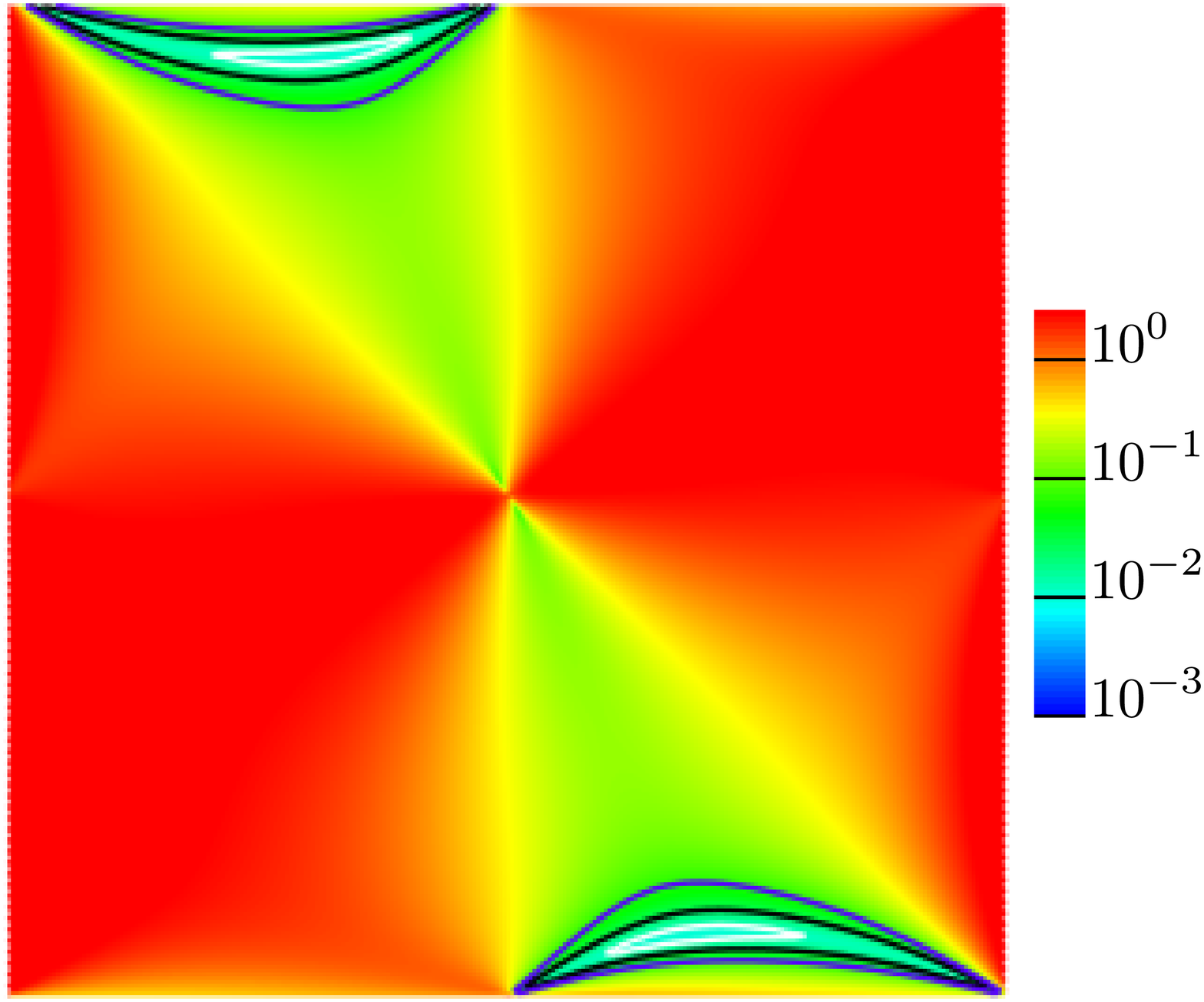}}
\caption{Color coded relative deviations Equ. (\ref{equ:relative_error}) of the displacement obtained numerically for three different vacancy configurations from the prediction of continuum theory as a function of $x$ and $y$. Color code and system size is the same as in Fig. (\ref{fig:reldev}). From top to bottom: $V_2$ vacancy ($\gamma=-1.78\sigma^2$), $V_3$ vacancy ($\gamma=-0.294\sigma^2$), and $V_a$ vacancy ($\gamma=-1.89\sigma^2$).}
\label{fig:reldevv2}
\end{figure}

In Fig. \ref{fig:reldevv2} we depict the relative deviations of the displacement of the particle simulation from the continuum theory calculated according to Equ. \ref{equ:relative_error}. Even far from the defect the displacement field obtained from the particle simulation does not become isotropic such that it cannot be reproduced by the continuum theory calculations. The best agreement is found for  $V_3$ (Fig. \ref{fig:reldevv2}, center), in which case the anisotropy of the displacement field is less  pronounced. The failure of the point defect model to reproduce the displacement patterns of vacancies, however, does not imply that elasticity theory is unsuitable for the description of such defects. Rather, the point defect model used here, which consists of two orthogonal pairs of opposing forces, appears to be to simple to capture the complex displacement patterns induced by vacancies.  

\section{Conclusion}
\label{sec:conclusion}

Interstitials and vacancies occur in various configurations generating displacement fields with symmetries that differ from the symmetry of the underlying triangular lattice. Near the defect, the displacement fields are highly anisotropic and strongly dependent on the atomistic details of the interactions. In this distance regime, linear elasticity theory brakes down due to discrete lattice effects and non-linearities of the potential. For distances larger than about 10-15 lattice spacings, however, elasticity theory is valid. To establish this validity, it is crucial that corresponding boundary conditions are used both in the continuum calculations and the particle-based numerical simulations. Since simulations are usually carried out with periodic boundary conditions in order to minimize finite size effects, the same boundary conditions must be employed also in the continuum calculation. If different boundary conditions are used, the long-range nature of elastic displacement fields can lead to considerable discrepancies even at length scale where elasticity theory is expected to hold.  

In this paper we have formulated the elastic theory problem in a way that makes it formally identical to the problem of determining the potential of a point charge in electrostatics. While here we have focused on two-dimensional systems, the same formalism applies also to three dimensions. Under periodic boundary conditions, this two-dimensional electrostatics problem has been solved using the method of Ewald summation \cite{LEEUW,NEUMANN}, in which the solution is expressed in terms of two rapidly convergent sums, one in real space and one in reciprocal space.

The solution of the electrostatics problem can be simply transferred to the continuum theory of the point defect. In this case, the role of the charge density in electrostatics is played by the dilatation, i.e. the local relative volume change. Accordingly, the charge neutrality required by the periodic boundary conditions in electrostatics corresponds to the condition of fixed volume in the elasticity theory. This requirement leads to a homogeneous neutralizing background that is seamlessly incorporated in the Ewald sum solution. The neutralizing background satisfies the condition of fixed volume by exactly compensating for the volume change caused by the introduction of the point defect. The volume compensation leads to an additional term in the energy density related to the work done by the defect against the external pressure. Depending on the pressure, this energy can contribute significantly to the total defect energy.  

While for interstitials the elasticity theory calculations carried out for a simple point defect model lead to good agreement with the particle calculations in the core region around the defect, large deviations are observed for vacancies. These discrepancies are due to the more complex displacement patterns of vacancies and better defect models are required to capture this behavior. 


\section*{Acknowledgments}
The authors would like to thank Christos Likos, Martin Neumann, and Andreas Tr\"oster and for useful discussions. This research was supported by the University of Vienna through the University Focus Research Area {\em Materials Science} (project ``Multi-scale Simulations of Materials Properties and Processes in Materials'').
 

\end{document}